\documentclass[preprint,superscriptaddress,aip,jcp,10pt]{revtex4-1}
\usepackage{graphicx,amsmath,amssymb,color,esint}

\newcommand{\rd}{\mathrm{d}}
\newcommand{\re}{\mathrm{e}}

\begin{document}

\title{Depolarized light scattering from prolate anisotropic particles: the influence of the particle shape on the field autocorrelation function}

\author{Christopher Passow}
\affiliation{Institut f{\"u}r Chemie, Universit{\"a}t Rostock, D-18051 Rostock, Germany}
\author{Borge ten Hagen}
\affiliation{Institut f{\"u}r Theoretische Physik II: Weiche Materie,
Heinrich-Heine-Universit{\"a}t D{\"u}sseldorf, D-40225 D{\"u}sseldorf, Germany}
\author{Hartmut L{\"o}wen}
\affiliation{Institut f{\"u}r Theoretische Physik II: Weiche Materie,
Heinrich-Heine-Universit{\"a}t D{\"u}sseldorf, D-40225 D{\"u}sseldorf, Germany}
\author{Joachim Wagner} 
\affiliation{Institut f{\"u}r Chemie, Universit{\"a}t Rostock, D-18051 Rostock, Germany}

\date{\today}

\begin{abstract}
We provide a theoretical analysis for the intermediate scattering function typically measured in  depolarized dynamic light scattering experiments. We calculate the field autocorrelation function $g_1^{\rm VH}(Q,t)$ in dependence on the wave vector $Q$ and the time $t$ explicitly in a vertical-horizontal scattering geometry for differently shaped solids of revolution. 
The shape of prolate cylinders, spherocylinders, spindles, and double cones with variable aspect ratio is expanded in rotational invariants $f_{lm}(r)$. By Fourier transform of these expansion coefficients, a formal multipole expansion of the scattering function is obtained, which is used to calculate the weighting coefficients appearing in the depolarized scattering function. In addition to translational and rotational diffusion, especially the translational-rotational coupling of shape-anisotropic objects is considered.
From the short-time behavior of the intermediate scattering function, the first cumulants $\Gamma(Q)$ are calculated. In a depolarized scattering experiment, they deviate from the simple proportionality to $Q^2$. The coefficients $f_{lm}(Q)$ strongly depend on the geometry and aspect ratio of the particles. The time dependence, in addition, is governed by the translational and rotational diffusion tensors, which are calculated by means of bead models for differently shaped particles in dependence on their aspect ratio. Therefore, our analysis shows how details of the particle shape---beyond their aspect ratio---can be determined by a precise scattering experiment.
This is of high relevance in understanding smart materials which involve suspensions of anisotropic colloidal particles. 
\end{abstract}


\maketitle

\section{Introduction}

\label{introduction}

Although the thermal motion of small suspended particles is connected to the name of Robert Brown, who observed a random motion of pollen in water with an optical microscope in 1827, this effect was for the first time reported by Jan Ingenhousz in 1785. More than  40 years earlier, Ingenhousz described the stochastic motion of inorganic charcoal dust on the surface of alcohol, already disproving Brown's hypothesis that this motion is caused by the vitality of pollen. The physical reason for this thermal motion was elucidated by Einstein in the annus mirabilis of physics, 1905.\cite{einstein:1905} Already one year later, he addressed the rotational diffusion of spherical particles.\cite{einstein:1906}

While the translational diffusion of colloidal particles has been thoroughly explored by experiments and theory,\cite{pusey:1991,Naegele:1996} rotational diffusion is much less investigated. The early starting point for the lasting interest in the diffusive motion of anisotropic particles is the well-known work of Perrin.\cite{perrin:1934} Experimental access to rotational diffusion is possible via dynamic light scattering employing laser light\cite{Berne:1973,King:1973,Aragon:1987,Degiorgio:1995,Semenov:1999,Pecora:2000,Glidden:2012,Alam:2014}
or coherent X-rays \cite{Holmqvist:2013}  as a probe. With near-field depolarized dynamic light scattering, in a heterodyne experiment, very small scattering vectors can be accessed.\cite{Brogioli:2009} Furthermore, birefringence \cite{Khlebtsov:2010} and NMR experiments \cite{Lee:1997} as well as computer simulations \cite{Loewen:1994,Kirchhoff:1996,Alvarez:2013} are used to investigate rotational diffusion. For particles with a size comparable to optical wavelengths, the rotational and translational motion can directly be observed by means of confocal video microscopy.\cite{Edmond:2012} As opposed to scattering methods, which immediately provide information on an ensemble average, video microscopy requires tracking a large number of trajectories for a statistical average. Furthermore, these methods are limited to comparatively large and slowly diffusing objects. 

While a variety of well-defined spherical particles have been studied, only few model systems of  well-defined anisotropic particles are available. Systems with tunable aspect ratio are even rarer. As model systems for prolate particles, e.g., goethite,\cite{lemaire:2002,lemaire:2004,lemaire:2004a,lemaire:2004b} hematite,\cite{ozaki:1984,maerkert:2011,Wagner:2013} polymer particles,\cite{Singh:2009} and viruses like tobacco mosaic viruses \cite{nemoto:1975,zasadzinski:1986,oldenbourg:1988,GrafPRE:1999} or fd viruses \cite{Dogic:1997,Lenstra:2001,Purdy:2003} have been described. With regard to oblate particles, gibbsite platelets are available  as a model system.\cite{wijnhoven:2005,wijnhoven:2005a}

Due to the orientation as an additional degree of freedom, however, the complexity of the phase diagram of such anisotropic particles is enhanced as compared to spherical particles since at least the interaction by the excluded volume is direction-dependent. Both prolate and oblate particles with large anisotropy are mesoscale analogs of liquid crystals \cite{vroege:1992,Bolhuis:1997} and therefore exhibit the same variety in phase behavior.
 
Anisotropic colloidal particles are important for many complex fluids in technical applications, e.g., suspended pigments in paints, clay particles in slurries used for the preparation of ceramics, and even in food engineering.\cite{Velikov:2008} 
Ferrofluids \cite{Odenbach:2004} and ferrogels \cite{Filipcsei:2007,Pessot:2014,Tarama:2014} 
(for a recent review see Ref.\ \citenum{Menzel:2015}) with magnetic particles provide further prominent examples of 
 smart materials where anisotropic particles are involved.
Both translational and rotational friction of those particles influence the macroscopic rheological behavior important
 for the processing of such fluids. Since the size of colloidal particles is similar to that of biological structures such as viruses or bacteria, their microrheology is also relevant to biological processes. 
 Shape-anisotropic particles, even when consisting of amorphous materials with isotropic physical properties, exhibit direction-dependent dielectric and magnetic properties and herewith related optical properties. Hence, the rotational motion of such particles affects their scattered intensity. 

The translational self-diffusion can be addressed experimentally by quasielastic scattering experiments using longitudinal and transversal waves without polarization analysis. For the experimental observation of rotational self-diffusion by means of scattering methods, however, transversal waves with polarization analysis are mandatory as a probe. The rotational motion can be elegantly
probed using an incident wave with a polarization perpendicular to that of the scattered wave. If the polarization of the incident beam is perpendicular to the scattering vector, the geometry of such a setup is called vertical-horizontal.

Let us consider an ensemble of freely rotating particles completely aligned along an initial unit vector $\boldsymbol{\Omega}_0$
at time $t=0$. This ensemble is described by a sharp orientational distribution function (odf) $p(\boldsymbol{\Omega})=\delta(\boldsymbol{\Omega}-\boldsymbol{\Omega}_0)$ with $\boldsymbol{\Omega}$ denoting a unit vector on the sphere.
Clearly, this orientational distribution function 
 becomes random with $p(\boldsymbol{\Omega})=1/(4\pi)$ after a very long time.
Hence, the mean angle between the initial and final orientation, $\left \langle\angle(\boldsymbol{\Omega}_0,\boldsymbol{\Omega}(t))\right\rangle_{t\to\infty}$, equals $\pi/2$, which corresponds to the angle enclosed by the  crossed polarizers in a depolarized scattering experiment.  

The key dynamical quantity measured in a depolarized light
scattering experiment in vertical-horizontal (VH) geometry is
the so-called intermediate scattering function, which is related to the field autocorrelation function
and provides access to the orientational dynamics. Even for a single Brownian particle, i.e., for a suspension at high dilution,
the time dependence of this correlation function is nontrivial due to the rotation-translation coupling in the laboratory frame.
A first theoretical approach to the translational-rotational coupling of cylinder-shaped objects is described by Arag\'on and Pecora \cite{Aragon:1985} using spheroidal basis functions. In this paper, we start from a systematic expansion of the particle form factors $P(Q,\boldsymbol{\Omega})$ into spherical harmonics. The intermediate scattering function in VH-geometry is then calculated by canonically averaging over rotational and translational transition probabilities obtained from the Smoluchowski equation for the Brownian dynamics of anisotropic particles.
 We give explicit expressions for the expansion coefficients and evaluate them for different
particle shapes such as cylinders, spherocylinders, ellipsoids, spindles, and double cones. 
We thereby show that details of the  particle shape---beyond the aspect ratio \cite{RodriguezFernandez:2007}---influence the static and dynamic scattering functions. Therefore, a precise scattering experiment, if evaluated within our framework, gives access to the details of the particle shape.
This is of high importance with regard to the interpretation of depolarized quasielastic scattering experiments employing coherent X-rays. When colloidal suspensions are studied, the size of the particles is significantly larger than X-ray wavelengths. Meanwhile, highly brilliant 
linearly polarized coherent X-rays are accessible from third generation synchrotron sources\cite{Blume:1988} and new upcoming free electron lasers.\cite{Suzuki:2014}

This paper is dedicated to the depolarized scattering of solids of revolution. Our approach can be applied to prolate as well as to oblate shapes. Here, at first prolate objects are considered.

The paper is organized as follows: in Sec.\ \ref{fieldautocorrelation}, we calculate the field autocorrelation function in VH-geometry in terms of rotational invariants. In section \ref{expansion}, the scattering length density of homogeneous solids of revolution is expanded in spherical harmonics. The static scattering functions are the Fourier transforms of these formal multipole expansions as described in Sec.\ \ref{scattering_function}. The method which is used to calculate the translational and rotational diffusion coefficients for the various considered particle shapes is presented in Sec.\ \ref{coefficients}. These diffusion coefficients are needed to describe the time dependence of the depolarized scattering function. The resulting first cumulants are analyzed in Sec.\ \ref{cumulants}, before we finally conclude in Sec.\ \ref{conclusion}.

\section{Depolarized field autocorrelation function $g_1^{\rm VH}(Q,t)$ for solids of revolution}
\label{fieldautocorrelation}

Let $p(\mathbf{r},\boldsymbol{\Omega},t)$ denote the probability density to find a particle at the time $t$ located with its center of mass at the position $\mathbf{r}$ and with orientation $\boldsymbol{\Omega}=(\sin \vartheta \cos \varphi, \sin \vartheta \sin \varphi, \cos \vartheta)$. The orientation $\boldsymbol{\Omega}$ is characterized by the angles $\{\vartheta,\varphi\}$, where $\vartheta$ is the polar angle and $\varphi$ the azimuthal one.

The translational and rotational diffusive motion of an ensemble of non-interacting particles can be described via the Smoluchowski equation\cite{Dhont_book}
\begin{align}
\left( \nabla_\mathbf{r}^{\rm T}\,.\, \left[ \begin{array}{ccc}D_\perp & & \\ & D_\perp & \\ & & D_\parallel \end{array} \right] \,.\, \nabla_\mathbf{r} + D_{\rm rot} \hat{\ell}^2\right) p(\mathbf{r},\boldsymbol{\Omega},t) &= \dfrac{\partial}{\partial t} p(\mathbf{r},\boldsymbol{\Omega},t)\,,
\end{align}
where $D_\parallel$ and $D_\perp$ are the translational diffusion coefficients parallel and perpendicular to the rotation axis of the body of revolution, while $D_{\rm rot}$ is the rotational diffusion coefficient for Brownian reorientations of the particle. $\nabla_\mathbf{r}$ is the spatial gradient, and the operator
\begin{align}
\hat{\ell}^2 = \dfrac{1}{\sin\vartheta}\left(\dfrac{\partial}{\partial\vartheta}\sin\vartheta\dfrac{\partial}{\partial\vartheta}\right)+\dfrac{1}{\sin^2\vartheta}\dfrac{\partial^2}{\partial\varphi^2}
\end{align}
is the angular part of the Laplacian acting on the orientation $\boldsymbol{\Omega}$ of the particle. 
 
In Fourier space, the Smoluchowski equation reads
\begin{align}
\left(-Q^2\; \hat{\mathbf{Q}}^{\rm T} . \left(\langle D_{\rm tr}\rangle \left[\begin{array}{ccc}1 & & \\ & 1 & \\ & & 1\end{array}\right] + \dfrac{1}{3}\,\Delta D_{\rm tr} \left[\begin{array}{ccc}-1 & & \\ & -1 & \\ & & 2\end{array}\right]\right). \hat{\mathbf{Q}} + D_{\rm rot}\hat{\ell}^2\right) p(\mathbf{Q},\boldsymbol{\Omega},t) &= \dfrac{\partial}{\partial t} p(\mathbf{Q},\boldsymbol{\Omega},t)\,,
\label{eq:FSmolu}
\end{align}
where $\mathbf{Q}$ is the scattering vector. The translational diffusion tensor is decomposed in irreducible spherical tensors $\mathbf{T}^{(0)}$ and
$\mathbf{T}^{(2)}$ using the abbreviations $\langle D_{\rm tr}\rangle = (2 D_\perp + D_\parallel)/3$ and 
$\Delta D_{\rm tr}=D_\parallel-D_\perp$. $\mathbf{T}^{(0)}$ is the unit tensor in the first term of Eq.\ \eqref{eq:FSmolu} and $\mathbf{T}^{(2)}$ is the traceless symmetric tensor in the second term, corresponding to formal monopole and quadrupole contributions to the translational diffusion. The contraction of the irreducible spherical tensors with the unit vectors $\hat{\mathbf{Q}}$ leads to
\begin{align}
\left(-Q^2 \langle D_{\rm tr}\rangle - \dfrac{2}{3}\,Q^2\Delta D_{\rm tr}\;P_2(\cos\vartheta)+ D_{\rm rot} \hat{\ell}^2\right)p(Q,\boldsymbol{\Omega},t)&=\dfrac{\partial}{\partial t}p(Q,\boldsymbol{\Omega},t)\,,
\end{align} 
where the polar angle $\vartheta$ corresponds to the angle enclosed between the particle axis assumed to be directed in the positive $z$-direction and the scattering vector $\mathbf{Q}=Q\boldsymbol{\Omega}$. Since scattering methods probe the projection of translational diffusive displacements on the scattering vector, the angle $\vartheta$ is also the angle enclosed between the particle's orientation and the direction of its diffusion path. $P_2(x)=(3x^2-1)/2$ is the second Legendre polynomial with $x=\cos\vartheta$. By means of the decomposition of the diffusion tensor in irreducible spherical tensors, the Smoluchowski equation in Fourier space for the translational and rotational diffusion of axially symmetric solids of revolution can formally be written as
\begin{align}
\label{eq:smoluchowski_L}
\hat{\mathcal{L}}p(Q,\boldsymbol{\Omega},t)&=\left(\hat{\mathcal{L}}_{\langle D_{\rm tr}\rangle}+\hat{\mathcal{L}}_{\Delta D_{\rm tr}}+\hat{\mathcal{L}}_{\rm rot}\right)p(Q,\boldsymbol{\Omega},t)=\dfrac{\partial}{\partial t}p(Q,\boldsymbol{\Omega},t)
\end{align}
with $\hat{\mathcal{L}}_{\langle D_{\rm tr}\rangle}=-Q^2 \langle D_{\rm tr}\rangle$ accounting for the isotropic or monopole part of the 
translational diffusion tensor, $\hat{\mathcal{L}}_{\Delta D_{\rm tr}}= -(2/3) Q^2\Delta D_{\rm tr} P_2(\cos\vartheta)$ for its quadrupole part, and, finally, $\hat{\mathcal{L}}_{\rm rot}=D_{\rm rot} \hat{\ell}^2$ for the rotational diffusion.

The solution of this parabolic partial differential equation is formally given by   
\begin{align}
p(Q,\boldsymbol{\Omega},t) = {\rm e}^{\hat{\mathcal{L}}t}\, p(Q,\boldsymbol{\Omega},0)\!\big|_{\boldsymbol{\Omega}=\boldsymbol{\Omega}_0} &=  {\rm e}^{\hat{\mathcal{L}}_{\langle D_{\rm tr}\rangle}t} {\rm e}^{\hat{\mathcal{L}}_{\Delta D_{\rm tr}}t} {\rm e}^{\hat{\mathcal{L}}_{\rm rot} t} p(Q,\boldsymbol{\Omega},0)\!\big|_{\boldsymbol{\Omega}=\boldsymbol{\Omega}_0} \,,
\end{align}
since the operators $\hat{\mathcal{L}}_{\langle D_{\rm tr}\rangle}$, $\hat{\mathcal{L}}_{\Delta D_{\rm tr}}$, and $\hat{\mathcal{L}}_{\rm rot}$ commute.

The depolarized scattering function $g_1^{\rm VH}(Q,t)$ is determined by the product of the solutions of the three parts of the Smoluchowski equation. As the different contributions are statistically independent, the scattering function can be written as
\begin{align}
\label{eq:prod}
g_1^{\rm VH}(Q,t) = g_{1,\langle D_{\rm tr} \rangle}^{\rm VH}(Q,t)  g_{1,\Delta D_{\rm tr}}^{\rm VH}(Q,t)  g_{1,{\rm rot}}^{\rm VH}(Q,t)\,.
\end{align}
Here, $g_{1,\langle D_{\rm tr} \rangle}^{\rm VH}(Q,t)$ accounts for the isotropic part of the translational diffusion, while $g_{1,\Delta D_{\rm tr}}^{\rm VH}(Q,t)$ gives the direction-dependent contribution. Finally, the effect of rotational diffusion is represented by the term $g_{1,{\rm rot}}^{\rm VH}(Q,t)$.
In the following, these parts are considered separately. Since for the calculation of the quadrupole contribution to the translational diffusion the results of both the rotational and the isotropic translational part are needed, the following two subsections briefly review corresponding textbook knowledge.\cite{berne:2000}

The depolarized field autocorrelation function is for an ergodic system given by \footnote{Note that the notation is slightly different than in Ref.\ \citenum{berne:2000}, where the conditional probability density $G(\boldsymbol{\Omega},t|\boldsymbol{\Omega}_0,0)$ is included in the definition of the angle brackets (see Eq.\ (7.3.12) on page 121 in Ref.\ \citenum{berne:2000}).}  
\begin{align}
\label{eq:field_autocorrelation}
g_1^{\rm VH}(Q,t) &= \dfrac{\left\langle\mathbf{E}^*(Q,\boldsymbol{\Omega}_0,0) G(\boldsymbol{\Omega},t|\boldsymbol{\Omega}_0,0)\mathbf{E}(Q,\boldsymbol{\Omega},t)\right\rangle_{\boldsymbol{\Omega}_0,\boldsymbol{\Omega}}}{\left\langle\mathbf{E}^*(Q,\boldsymbol{\Omega}_0,0)\mathbf{E}^*(Q,\boldsymbol{\Omega}_0,0)\right\rangle_{\boldsymbol{\Omega}_0}}\,,
\end{align}
where the notation $\langle \ldots \rangle_{\boldsymbol{\Omega}_0,\boldsymbol{\Omega}}$ indicates canonical averaging over  the initial orientation $\boldsymbol{\Omega}_0$ and the final orientation $\boldsymbol{\Omega}$ of the particles. The scattered electromagnetic field expanded in spherical harmonics reads 
\begin{align}
\label{eq:field_expansion}
\mathbf{E}(Q,\boldsymbol{\Omega}(t))=\sum\limits_l\sum\limits_m f_{lm}(Q)Y_{lm}(\boldsymbol{\Omega}(t))
\end{align}
and $\mathbf{E}^*(Q,\boldsymbol{\Omega}(t))$ is its complex conjugate.
The kernel $G(\boldsymbol{\Omega},t|\boldsymbol{\Omega}_0,0)$ is the conditional probability density that an object's orientation is $\boldsymbol{\Omega}$ at time $t$ given that it was $\boldsymbol{\Omega}_0$ at the time $t_0=0$. This conditional probability density is the product of the transition probability $p(\boldsymbol{\Omega},t|\boldsymbol{\Omega}_0,0)$ for a rotation from $\boldsymbol{\Omega}_0$ to $\boldsymbol{\Omega}$ and the probability density to find the particle in a statistical ensemble with the initial orientation $\boldsymbol{\Omega}_0$:
\begin{align}
\label{eq:transition_probability}
G(\boldsymbol{\Omega},t|\boldsymbol{\Omega}_0,0)& =\dfrac{1}{4\pi} p(\boldsymbol{\Omega},t|\boldsymbol{\Omega}_0,0)\,.
\end{align}

\subsection{Depolarized field autocorrelation function for rotational diffusion $g_{1,\rm rot}^{\rm VH}(Q,t)$}
In a depolarized light scattering experiment, the sample is illuminated by linearly polarized light with the electric field vector oriented in $z$-direction.
Both the scattering vector $\mathbf{Q}$ and the electric field vector of the scattered light are perpendicular to the $z$- direction. 
In VH-geometry, scattered light is only detected if the electric field vector changes its orientation during the scattering process by a rotation of the scattering objects. 
To calculate the statistical average over the initial orientation $\boldsymbol{\Omega}_0$ and the final orientation $\boldsymbol{\Omega}$, the transition probability $p(\boldsymbol{\Omega},t|\boldsymbol{\Omega}_0,0)$ between these two orientations is needed. For a freely diffusing particle it can be obtained from the rotational part 
\begin{align}
\label{eq:rot}
\hat{\mathcal{L}}_{\rm rot} p(\boldsymbol{\Omega},t|\boldsymbol{\Omega}_0,0) = D_{\rm rot} \hat{\ell}^2 p(\boldsymbol{\Omega},t|\boldsymbol{\Omega}_0,0) = \dfrac{\partial}{\partial t} p(\boldsymbol{\Omega},t|\boldsymbol{\Omega}_0,0)
\end{align}
of the Smoluchowski equation (\ref{eq:smoluchowski_L}).
The formal solution of Eq.\ \eqref{eq:rot} is
\begin{align}
p(\boldsymbol{\Omega},t|\boldsymbol{\Omega}_0,0)&=\sum\limits_{l=0}^\infty\sum\limits_{m=-l}^l \exp\left(-D_{\rm rot}t\hat{\ell}^2\right)Y_{lm}^*(\boldsymbol{\Omega})Y_{lm}(\boldsymbol{\Omega}_0) \nonumber \\
&=\sum\limits_{l=0}^\infty \sum\limits_{m=-l}^{l} \exp\left(-l(l+1)D_{\rm rot}t\right) Y_{lm}^*(\boldsymbol{\Omega})Y_{lm}(\boldsymbol{\Omega}_0)\,.
\end{align}

\begin{widetext}
Combining Eqs.\ \eqref{eq:field_autocorrelation}, \eqref{eq:field_expansion}, and \eqref{eq:transition_probability} leads to the expression
\begin{align}
g_{1,\rm rot}^{\rm VH}(Q,t) &= \dfrac{\bigg\langle \sum\limits_{l_1m_1} f_{l_1m_1}(Q) Y_{l_1m_1}^*(\boldsymbol{\Omega}_0)\;G(\boldsymbol{\Omega},t|\boldsymbol{\Omega}_0,0)\; \sum\limits_{l_2m_2} f_{l_2m_2}(Q)Y_{l_2m_2}(\boldsymbol{\Omega})\bigg\rangle_{\boldsymbol{\Omega}_0,\boldsymbol{\Omega}}}{\bigg\langle \sum\limits_{l_1m_1} f_{l_1m_1}(Q) Y_{l_1m_1}^*(\boldsymbol{\Omega}_0)\: p(\boldsymbol{\Omega}_0)\: \sum\limits_{l_2m_2} f_{l_2m_2}(Q)Y_{l_2m_2}(\boldsymbol{\Omega}_0)\bigg\rangle_{\boldsymbol{\Omega}_0}} \nonumber \\
&=\dfrac{\oiint\limits_{4\pi}\oiint\limits_{4\pi}
\sum\limits_{lm}f_{lm}(Q)Y_{lm}^*(\boldsymbol{\Omega}_0)\sum\limits_{l'm'} \exp\left(-l(l+1)D_{\rm rot}t\right) Y_{l'm'}(\boldsymbol{\Omega}_0)Y_{l'm'}^*(\boldsymbol{\Omega}) \sum\limits_{l''m''}f_{l''m''}(Q)Y_{l''m''}(\boldsymbol{\Omega})  \rd\boldsymbol{\Omega}\,\rd\boldsymbol{\Omega}_0
}
{\oiint\limits_{4\pi}\sum\limits_{lm}f_{lm}(Q)Y_{lm}^*(\boldsymbol{\Omega}_0)\sum\limits_{l'm'}f_{l'm'}(Q)Y_{l'm'}(\boldsymbol{\Omega}_0)\;\rd \boldsymbol{\Omega}_0} 
\end{align}
for the field autocorrelation function. Here, $\rd \boldsymbol{\Omega}=\sin \vartheta \, \rd \vartheta \, \rd \varphi$  is the differential for the orientational integrals.
\end{widetext}
With the orthogonality relation
\begin{align}
\oiint\limits_{4\pi} Y_{lm}^*(\boldsymbol{\Omega})Y_{l'm'}(\boldsymbol{\Omega})\,\rd\boldsymbol{\Omega}=\delta_{ll'}\delta_{mm'}\label{eq:orthogonality}
 \end{align}
  one obtains
\begin{align}
\label{eq:autocorrelation_rot}
g_{1,\rm rot}^{\rm VH}(Q,t)  &=\dfrac{ \sum\limits_{lm} \sum\limits_{l'm'} \sum\limits_{l''m''}
f_{lm}(Q) \exp\left(-l(l+1)D_{\rm rot}t\right) \delta_{ll'}\delta_{mm'} f_{l''m''}(Q) \delta_{l'l''}\delta_{m'm''}} {\sum\limits_{lm}\sum\limits_{l'm'} f_{lm}(Q)f_{l'm'}(Q) \delta_{ll'}\delta_{mm'}} \nonumber \\ 
&=\dfrac{\sum\limits_{lm} f_{lm}^2(Q)\exp\left(-l(l+1)D_{\rm rot}t\right)}{\sum\limits_{lm}f_{lm}^2(Q)}=\dfrac{1}{\langle P(Q,\boldsymbol{\Omega})\rangle_{\boldsymbol{\Omega}}}\sum\limits_{lm} f_{lm}^2(Q)\exp\left(-l(l+1)D_{\rm rot}t\right).
\end{align}
The orientation-averaged particle form factor
\begin{align}
P(Q)=\left\langle P(Q,\boldsymbol{\Omega})\right\rangle_{\boldsymbol{\Omega}}=\sum_{lm}f_{lm}^2(Q)
\end{align}
is the scattered intensity of a randomly aligned ensemble normalized to the forward scattering, i.~e., $\lim_{Q\to 0} P(Q)\equiv 1$.

\subsection{Depolarized field autocorrelation function for translational diffusion: the isotropic part $g_{1,\langle D_{\rm tr}\rangle}^{\rm VH}(Q,t)$}
By means of the short-time limit of the orientational transition probability density
\begin{align}
\lim\limits_{t\to 0} G(\boldsymbol{\Omega},t | \boldsymbol{\Omega}_0,0) = \dfrac{1}{4\pi}\sum\limits_{lm} Y^*_{lm}(\boldsymbol{\Omega})Y_{lm}(\boldsymbol{\Omega}_0) = \frac{1}{4\pi}\delta(\boldsymbol{\Omega}-\boldsymbol{\Omega}_0)
\label{eq:Glim}
\end{align}
the contribution of translational diffusion to the field autocorrelation function is accessible. Physically, the limit $t \to 0$ means that the orientational degrees of freedom are frozen.
The initial condition $p(\mathbf{r},0)=\delta(\mathbf{r})$ for translational diffusion reads $p(\mathbf{Q},0)=1$ in Fourier space.
Hence, with the notation $\mathbf{Q}=Q\boldsymbol{\Omega}$ the solution of the partial differential equation 
\begin{align}
\hat{\mathcal{L}}_{\langle D_{\rm tr}\rangle} p(Q,\boldsymbol{\Omega},t)&=-Q^2\langle D_{\rm tr}\rangle p(Q,\boldsymbol{\Omega},t) =\dfrac{\partial}{\partial t}p(Q,\boldsymbol{\Omega},t)
\end{align}
can be written as
\begin{align}
p(Q,\boldsymbol{\Omega},t) &= \exp\left(-Q^2 \langle D_{\rm tr}\rangle t\right)\,,
\end{align} 
since a scattering experiment selectively probes the projection of the diffusive path to the scattering vector $\mathbf{Q}$. 

Using Eq.\ \eqref{eq:Glim}, the depolarized field autocorrelation function for the orientation-averaged translational diffusion of an ensemble of randomly aligned particles reads
\begin{widetext}
\begin{align}
g_{1,\langle D_{\rm tr}\rangle}^{\rm VH}(Q,t) &=\dfrac{
\left\langle\sum\limits_{lm}f_{lm}(Q)Y_{lm}^*(\boldsymbol{\Omega}_0)\;\frac{1}{4\pi}\delta(\boldsymbol{\Omega}-\boldsymbol{\Omega}_0)\;\exp\left(-Q^2\langle D_{\rm tr}\rangle t\right) 
\sum\limits_{l'm'}f_{l'm'}(Q)Y_{l'm'}(\boldsymbol{\Omega})\right\rangle_{\boldsymbol{\Omega}_0,\boldsymbol{\Omega}}}
{\left\langle\sum\limits_{lm}f_{lm}(Q)Y_{lm}^*(\boldsymbol{\Omega}_0)\;p(\boldsymbol{\Omega}_0)\; 
\sum\limits_{l'm'}f_{l'm'}(Q)Y_{l'm'}(\boldsymbol{\Omega}_0)\right\rangle_{\boldsymbol{\Omega}_0}} \,.
\label{eq:g1Dtr}
\end{align}
\end{widetext}

After performing one orientational integration, the numerator and the denominator in Eq.\ \eqref{eq:g1Dtr} are identical except for the exponential factor. Thus, one obtains 
\begin{align}
\label{eq:autocorrelation_daverage}
g_{1,\langle D_{\rm tr}\rangle}^{\rm VH}(Q,t)  & = \exp\left(-Q^2\langle D_{\rm tr}\rangle t \right).
\end{align}
This contribution is the only one also visible in polarized dynamic light scattering experiments, where the polarization of both incident and scattered beam is vertical (VV-geometry). As a consequence,  such experiments are only sensitive to the orientation-averaged translational diffusion coefficient.

\subsection{Depolarized field autocorrelation function for translational diffusion: the anisotropic part $g_{1,\Delta D_{\rm tr}}^{\rm VH}(Q,t)$}

 The final contribution $g_{1,\Delta D_{\rm tr}}^{\rm VH}(Q,t)$ to the depolarized field autocorrelation function originates from the anisotropy of the translational diffusion.  This contribution has only been addressed for infinitely thin rods so far.\cite{Dhont_book} Here, we provide a general expression in terms of rotational invariants. The partial differential equation 
\begin{align}
\hat{\mathcal{L}}_{\Delta D_{\rm tr}} p(Q,\boldsymbol{\Omega},t) = -\dfrac{2}{3}Q^2\Delta D_{\rm tr}\,P_2(\cos\vartheta) p(Q,\boldsymbol{\Omega},t) = \dfrac{\partial}{\partial t} p(Q,\boldsymbol{\Omega},t)
\end{align} 
resulting from the traceless symmetric tensor $\mathbf{T}^{(2)}$ has the formal solution
\begin{align}
p(Q,\boldsymbol{\Omega},t) = \exp\left(-\dfrac{2}{3}Q^2\Delta D_{\rm tr} P_2(\cos\vartheta) t\right)\,. 
\end{align}
The second Legendre polynomial can be written as $P_2(\cos\vartheta)=(4\pi/5)^{1/2}Y_{2,0}(\boldsymbol{\Omega})$ in terms of spherical harmonics.

Herewith, the contribution of the depolarized scattering function related to the anisotropy of the translational diffusion is given by
\begin{align}
g_{1,\Delta D_{\rm tr}}^{\rm VH}(Q,t) &=\dfrac{
\left\langle\sum\limits_{lm}f_{lm}(Q)Y_{lm}^*(\boldsymbol{\Omega}_0)\frac{1}{4\pi}\delta(\boldsymbol{\Omega}-\boldsymbol{\Omega}_0)
\exp\left(-\dfrac{2}{3}Q^2\Delta D_{\rm tr} \left(\dfrac{4\pi}{5}\right)^\frac{1}{2}Y_{2,0}(\boldsymbol{\Omega})t\right)\sum\limits_{l'm'}(Q)Y_{l'm'}(\boldsymbol{\Omega}) \right\rangle_{\boldsymbol{\Omega}_0,\boldsymbol{\Omega}}
}
{\left\langle\sum\limits_{lm}f_{lm}(Q)Y_{lm}^*(\boldsymbol{\Omega}_0)\;p(\boldsymbol{\Omega}_0)\; 
\sum\limits_{l'm'}f_{l'm'}(Q)Y_{l'm'}(\boldsymbol{\Omega}_0)\right\rangle_{\boldsymbol{\Omega}_0}}\,.
\end{align}
The short-time behavior of $g_{1,\Delta D_{\rm tr}}^{\rm VH}(Q,t)$ is obtained by expanding the exponential function in 
a Taylor series. In the short-time limit we obtain
\begin{align}
\label{eq:Taylor_expansion}
p(Q,\boldsymbol{\Omega},t) &= 1-\dfrac{2}{3}Q^2\Delta D_{\rm tr} \left(\dfrac{4\pi}{5}\right)^\frac{1}{2}Y_{2,0}(\boldsymbol{\Omega}) t + \mathcal{O}(t^2)\,.
\end{align}
\begin{widetext}
With this expansion the depolarized scattering function $g_{1,\Delta D_{\rm tr}}^{\rm VH}(Q,t)$ reads
\begin{align}
g_{1,\Delta D_{\rm tr}}^{\rm VH}(Q,t) &=\dfrac{
\oiint\limits_{4\pi} \sum\limits_{lm}f_{lm}(Q)Y_{lm}^*(\boldsymbol{\Omega}_0)\sum\limits_{l'm'}f_{l'm'}(Q) Y_{l'm'}(\boldsymbol{\Omega}_0)\,\rd\boldsymbol{\Omega}_0
}
{\oiint\limits_{4\pi}\sum\limits_{lm}f_{lm}(Q)Y_{lm}^*(\boldsymbol{\Omega}_0)\sum\limits_{l'm'}f_{l'm'}(Q)Y_{l'm'}(\boldsymbol{\Omega}_0)\,\rd\boldsymbol{\Omega}_0
} \nonumber \\
&\quad - \dfrac{\dfrac{2}{3}Q^2\Delta D_{\rm tr}\,t\left(\dfrac{4\pi}{5}\right)^\frac{1}{2}\oiint\limits_{4\pi}
\sum\limits_{lm}f_{lm}(Q)Y_{lm}^*(\boldsymbol{\Omega}_0) Y_{2,0}(\boldsymbol{\Omega}_0)\sum\limits_{l'm'}f_{l'm'}(Q)Y_{l'm'}(\boldsymbol{\Omega}_0)
 \,\rd\boldsymbol{\Omega}_0}
{\oiint\limits_{4\pi}\sum\limits_{lm}f_{lm}(Q)Y_{lm}^*(\boldsymbol{\Omega}_0)\sum\limits_{l'm'}f_{l'm'}(Q)Y_{l'm'}(\boldsymbol{\Omega}_0)\,\rd\boldsymbol{\Omega}_0} +\mathcal{O}(t^2)\,,
\end{align}
where the integration with respect to $\boldsymbol{\Omega}$ has already been performed.  
Using the orthogonality relation \eqref{eq:orthogonality}, this leads to 
\begin{align}
g_{1,\Delta D_{\rm tr}}^{\rm VH}(Q,t)&=
 1- \dfrac{2Q^2\Delta D_{\rm tr}t}{3\,P(Q)}\left(\dfrac{4\pi}{5}\right)^\frac{1}{2}\sum\limits_{lm}\sum\limits_{l'm'} f_{lm}(Q) \oiint\limits_{4\pi}
Y_{lm}^*(\boldsymbol{\Omega}_0)Y_{2,0}(\boldsymbol{\Omega}_0) f_{l'm'}(Q)Y_{l'm'}(\boldsymbol{\Omega}_0)
\,\rd\boldsymbol{\Omega}_0+\mathcal{O}(t^2)\,.
\end{align}
\end{widetext}
The remaining integral over three spherical harmonics depending on $\boldsymbol{\Omega}_0$ can be solved employing
Gaunt's theorem.\cite{gray:1984} Thus, one finally obtains
\begin{align}
\label{eq:autocorrelation_DeltaD}
g_{1,\Delta D_{\rm tr}}^{\rm VH}(Q,t)&=1-\dfrac{2Q^2\Delta D_{\rm tr}\,t}{3\,P(Q)}\left(\dfrac{4\pi}{5}\right)^\frac{1}{2}\sum\limits_{lm}\sum\limits_{l'm'} f_{lm}(Q)f_{l'm'}(Q) \nonumber \\
&\qquad\times \left(\dfrac{5 (2l'+1)}{4\pi (2l+1)}\right)^\frac{1}{2} C(l',2,l;m',0,m) C(l',2,l;0,0,0) +\mathcal{O}(t^2)\nonumber \\ \nonumber
&=1-\dfrac{2 Q^2\Delta D_{\rm tr}\,t}{3 P(Q)}\sum\limits_{lm}\sum\limits_{l'm'} f_{lm}(Q) f_{l'm'}(Q)\\ 
&\qquad \times \left(\dfrac{2l'+1}{2l+1}\right)^\frac{1}{2}C(l',2,l;m',0,m)C(l',2,l;0,0,0)+\mathcal{O}(t^2)
\end{align}
with the Clebsch-Gordan coefficients $C(l_1,l_2,l_3; m_1,m_2,m_3)$.

For solids of revolution with an equatorial mirror plane, only the coefficients $f_{l,0}$ with even $l$ differ from zero.
Hence, in this case just contributions 
\begin{align}
C(Q)&=\dfrac{2}{3 P(Q)}\sum\limits_{l}\sum\limits_{l'}
\left(\dfrac{4l'+1}{4l+1}\right)^{1/2}C^2(2l',2,2l;0,0,0)f_{2l,0}(Q)f_{2l',0}(Q) 
\end{align}
have to be considered. Only Clebsch-Gordan coefficients $C(2l',2,2l;0,0,0)$ with $|2l'-2l|\le 2$ are nonzero.
The relevant Gaunt coefficients appearing in the coupling function of solids of revolution with an equatorial mirror plane 
are
\begin{align}
c_{l,l'}&=\left(\dfrac{4l'+1}{4l+1}\right)^{1/2}C^2(2l',2,2l;0,0,0)\,.
\end{align}
With these coefficients the function $C(Q)$ can be written as
\begin{align}
\label{eq:coupling_function}
C(Q)&=\dfrac{2}{3 P(Q)}\sum\limits_{lm}\sum\limits_{l'm'} c_{l,l'} f_{lm}(Q) f_{l'm'}(Q)
\end{align}
leading to the expression 
\begin{align}
g_{1,\Delta D_{\rm tr}}^{\rm VH}(Q,t)&=1- C(Q) Q^2 \Delta D_{\rm tr} t + \mathcal{O}(t^2) 
\end{align}
for the short-time behavior of the field autocorrelation function resulting from the anisotropy of the translational diffusion. Iteration for a finite time $t$ then yields
\begin{align}
\label{eq:autocorrelation_danisotropic}
 g_{1,\Delta D_{\rm tr}}^{\rm VH}(Q,t)&= \exp\left(-C(Q) Q^2 \Delta D_{\rm tr}\,t\right) .
\end{align}
Unlike the weighting factors for the rotational contribution [see Eq.\ \eqref{eq:autocorrelation_rot}], the coupling function can be calculated via a different route avoiding the expansion in spherical harmonics \cite{Dhont_book} as the orientation average of the form factor weighted by $2P_2(\cos\vartheta)/3$.  
If $P(Q,\vartheta)$ denotes the form factor of a particle with the orientation $\boldsymbol{\Omega}$, the coupling function $C(Q)$ for solids of revolution can be written as 
\begin{align}
C(Q) &= \dfrac{1}{\langle P(Q,\vartheta)\rangle_{\vartheta}}\int\limits_{0}^\pi P(Q,\vartheta)  \dfrac{2}{3}P_2(\cos\vartheta)\,\sin\vartheta \,\rd \vartheta\,,
\end{align}
since the angular dependence of the diffusive quadrupole contribution $(l=2)$ just leads to a factor $2P_2(\cos\vartheta)/3$. In the rotational part, however, the multipole contributions are weighted by $l(l+1)$, which makes a separation of the multipole contributions necessary.

In the context of scattering from solids of revolution, the term translational-rotational coupling denotes a dependence of the translational mobility and the orientation of the particle in the laboratory coordinate system defined by the scattering vector: if a prolate particle has the orientation $\boldsymbol{\Omega}$ at time $t$, its mobility parallel to the orientation vector is larger than perpendicular. Since scattering experiments, however, probe the projection of the diffusive path to the scattering vector, also the rotation of anisotropic particles influences the scattered intensity related to translational diffusion.

As a side remark, we note that for objects of more complex shape, where due to the absence of inversion symmetry nonzero off-diagonal elements in the diffusion tensor exist, in terms of hydrodynamics another kind of translational-rotational coupling occurs.\cite{Wittkowski:2012,Kraft:13} In that case, a rotation causes a translation and vice versa, leading to a translational-rotational coupling even in the particle's frame of reference. Such effects, however, are not present for shapes with inversion symmetry.

\section{Expansion of the scattering function in spherical harmonics}
\label{expansion}

For the calculation of the depolarized scattering functions in Eqs.\ \eqref{eq:autocorrelation_rot} and  \eqref{eq:autocorrelation_DeltaD} the expansion coefficients $f_{lm}(Q)$ are required. These quantities are Hankel transforms of the expansion coefficients $f_{lm}(r)$. The latter quantities are calculated in this section.

Let us define the reduced scattering length density of a solid of revolution in terms of polar coordinates 
\begin{align}
\Psi(r,\vartheta,\varphi) &=\left\{\begin{array}{cll} 1&:&\mbox{inside}\\ 0 &:&\mbox{outside}\end{array} \right.
\end{align}
of the particle. Assuming axial symmetry in $z$-direction, this function is independent of
the  azimuthal angle $\varphi$, hence $\Psi(r,\vartheta,\varphi)\equiv\Psi(r,\vartheta)$. Let us additionally assume inversion symmetry related to the particle's center of mass. Then, together with the axial symmetry, an equatorial mirror plane is generated leading to the parity relation
\begin{align}
\Psi(r,\vartheta)&=\Psi(r,\pi-\vartheta)\,.
\end{align}
The spherical harmonics $Y_{lm}(\boldsymbol{\Omega})$ with
\begin{align}
Y_{lm}(\vartheta,\varphi) &= \left(\dfrac{(2l+1)}{4\pi}\dfrac{(l-m)!}{(l+m)!}\right)^{\frac{1}{2}} P_{lm}(\cos \vartheta) \re^{\imath m \varphi}\quad\forall\, m\ge 0 \quad \mbox{and} \\
Y_{lm}(\vartheta,\varphi) &= (-1)^{-m} Y_{l,-m}^*(\boldsymbol{\Omega})\quad \forall\, m<0
\end{align}
form a complete orthogonal basis. Hence, the reduced scattering length density can be expanded in spherical harmonics
\begin{align}
\Psi(r,\vartheta)&=\sum\limits_{l=0}^\infty\sum\limits_{m=-l}^m f_{lm}(r)\,Y_{lm}(\vartheta,\varphi)
\end{align}
with the expansion coefficients $f_{lm}(r)$ given by
\begin{align}
\label{eq:flm}
f_{lm}(r)&=\oiint\limits_{4\pi} Y_{lm}^*(\boldsymbol{\Omega})\,\Psi(r,\boldsymbol{\Omega})\,\rd\boldsymbol{\Omega}\,.
\end{align}

\subsection{Intersections of solids of revolution with spheres}

\begin{figure}[tb]
\centerline{\includegraphics[height=50mm]{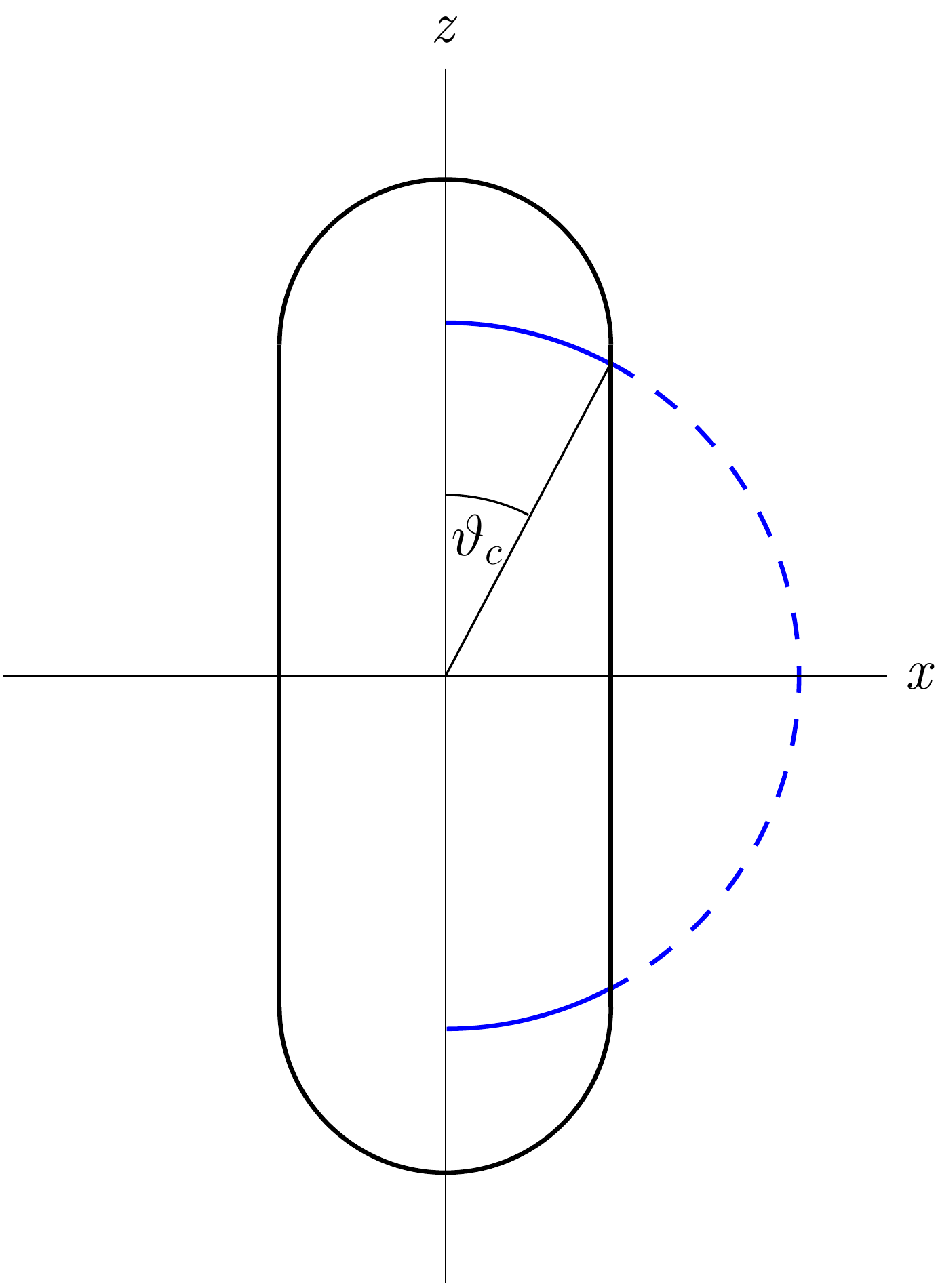}\hspace*{5mm}\includegraphics[height=50mm]{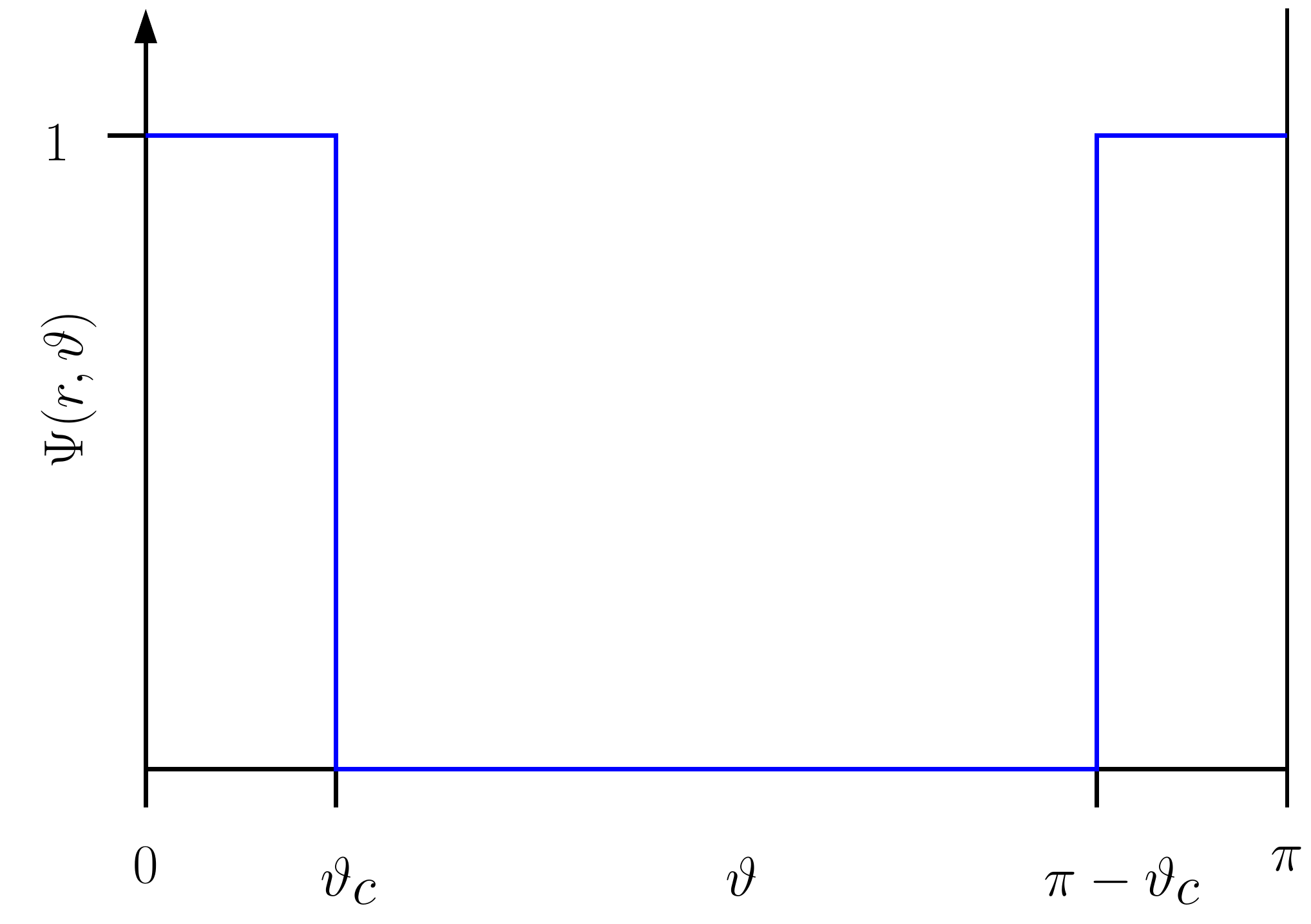}}
\caption{\label{fig:weighting_function}The reduced scattering length density $\Psi(r,\vartheta)$ as a step function of the critical pole distances $\vartheta_c$ and $\pi-\vartheta_c$ exemplarily shown for spherocylinders. The critical pole distances $\vartheta_c$ and $\pi-\vartheta_c$ depending on the radius $r$  are accessible from the meridian curve $x(z)$ of the particles.}
\end{figure}

With knowledge of the meridian curve $x(z)$ of a solid of revolution, the function $\Psi(r,\vartheta)$ can easily be obtained (see Fig.\ \ref{fig:weighting_function}), if the polar angle $\vartheta_c(r)$ of the intersection of a sphere with the radius $r$ and the solid of revolution is calculated. In dependence on $\vartheta_c(r)$ the function $\Psi(r,\vartheta)$
can be written as 
\begin{align}
\Psi(r,\vartheta_c)&=1-\mathcal{H}(\cos\vartheta_c)+\mathcal{H}(\cos(\pi-\vartheta_c))\,,
\end{align}
where $\mathcal{H}(\cos\vartheta)$ denotes the Heaviside function
\begin{align}
\mathcal{H}(\cos\vartheta_c)=\int\limits_{-1}^1 \delta(\cos(\vartheta-\vartheta_c))\,\rd\cos\vartheta \,.
\end{align}

For the calculation of the expansion coefficients the pole distances of intersections of spheres and the solids of revolution are necessary. Let $\xi=\cos\vartheta_c$ be the cosine of the critical pole distance where the sphere intersects the solid of revolution. Due to the equatorial mirror plane, an identical intersection can be found at $\pi-\vartheta_c$. The critical polar angles are accessible via the meridian functions $x(z)$.  Let us for simplicity use reduced units, i.e., we set the equatorial radius to $R_{\rm eq}=1$.

\begin{figure}[tb]
\centerline{\includegraphics[height=80mm]{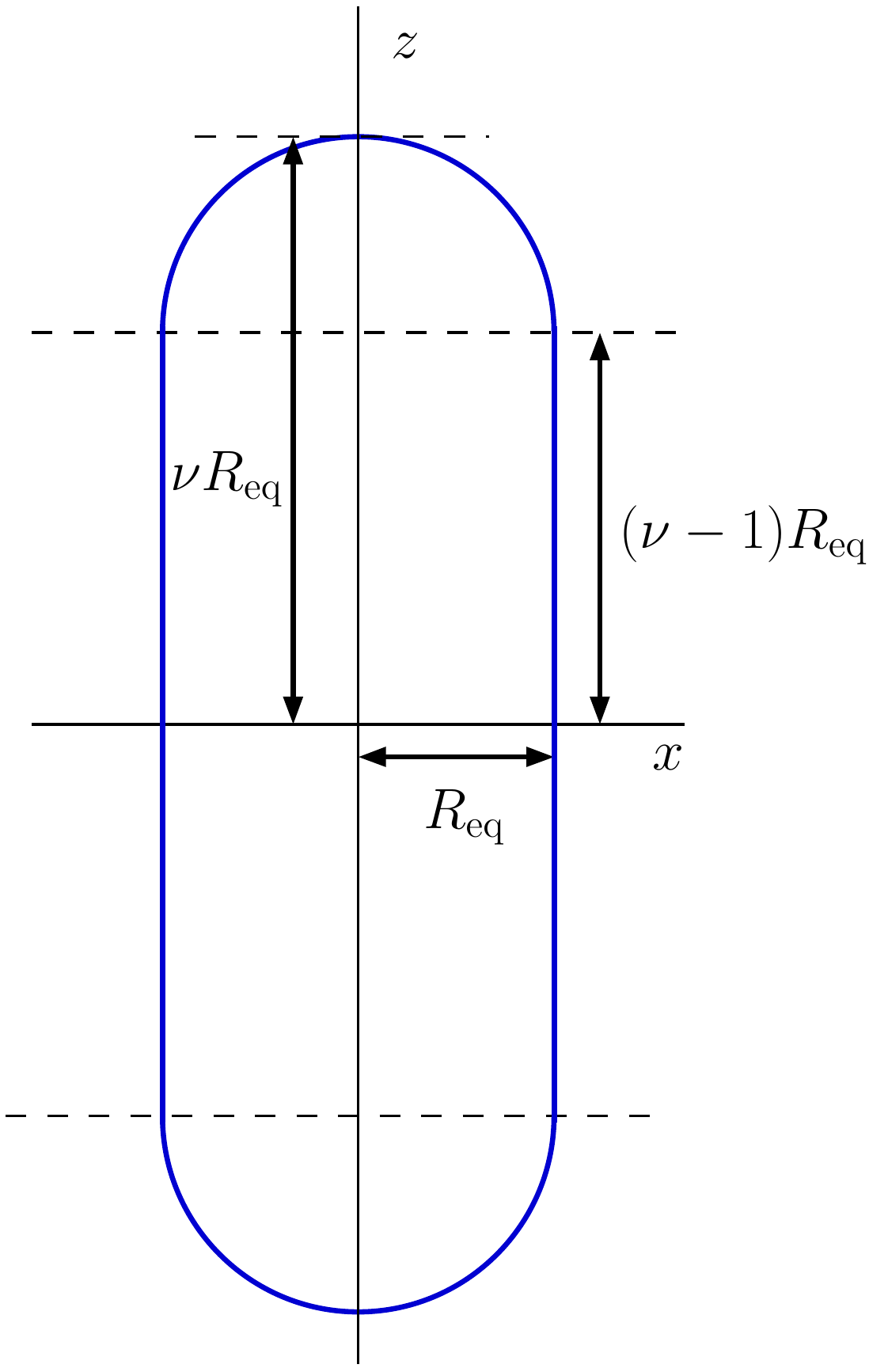}\hspace*{5mm} \includegraphics[height=80mm]{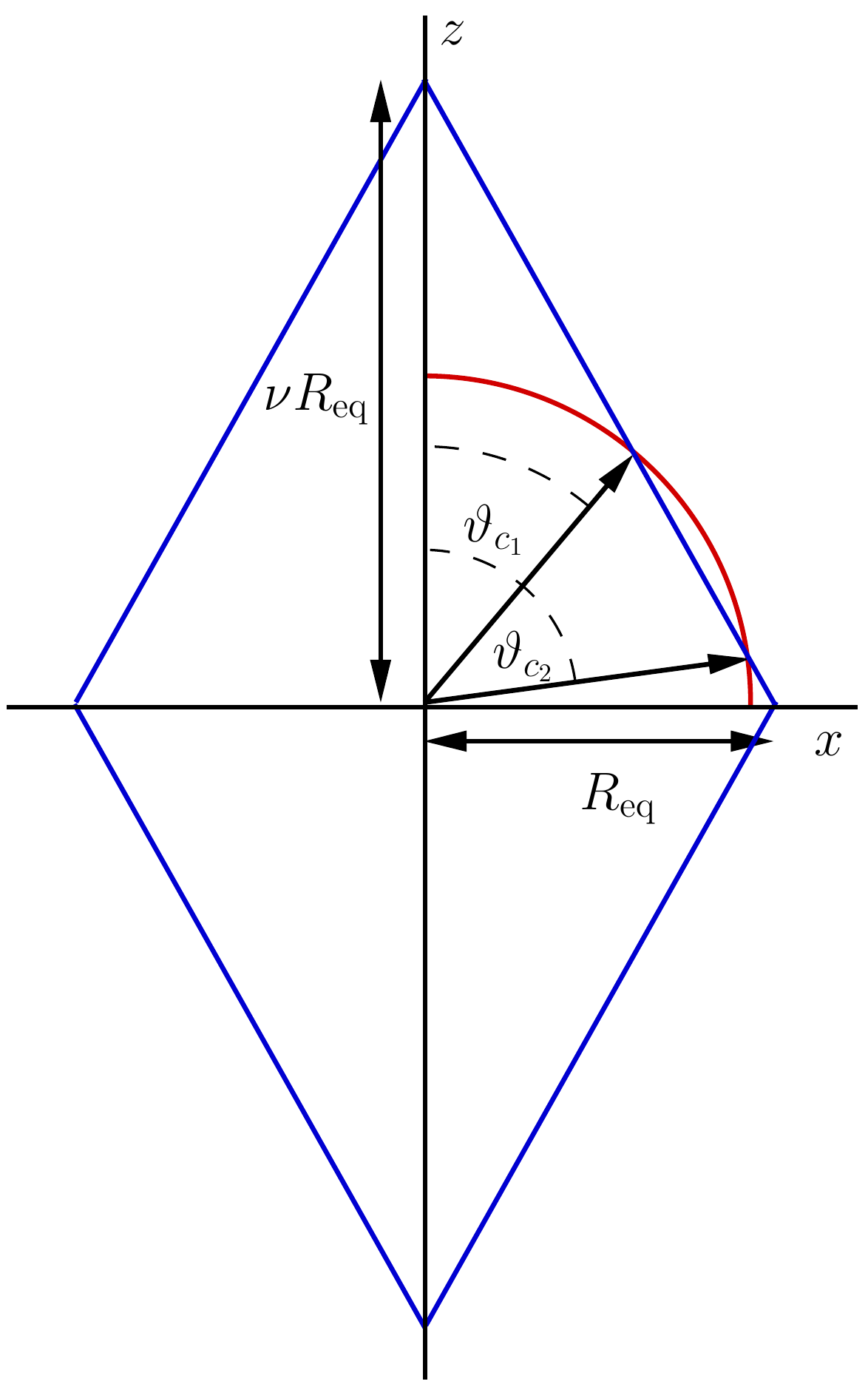}}
\caption{\label{fig:critical_angles}Meridian curves for spherocylinders (lhs) and double cones (rhs) as well as critical pole distances for intersections with spheres.}
\end{figure}

Exemplarily, we discuss the expansion for spherocylinders and double cones here, while cylinders, ellipsoids, and spindles are addressed in Appendix A. A spherocylinder  with the aspect ratio $\nu$ consists of a cylinder with the radius $R_{\rm eq}$ and the length $2R_{\rm eq}(\nu-1)$ with two hemispheres with centers at $(\nu-1)R_{\rm eq}$ and $-(\nu-1)R_{\rm eq}$ (see Fig.\ \ref{fig:critical_angles}).  

Hence, the meridian curve for a spherocylinder with the rotation axis in $z$-direction is
\begin{align}
x(z)&=\left\{\begin{array}{lcl}1&:& -(\nu-1)\le z\le \nu-1 \\ \left(1-(z-(\nu-1))^2\right)^{1/2}&:& |z|>\nu-1\end{array}\right. \,.
\end{align}
The symmetrically equivalent pole distances of intersections with spheres can be written as
\begin{align}
\xi=\cos\vartheta_c&=\left\{\begin{array}{lcl}\pm\left(\dfrac{r^2-1}{r^2}\right)^{\frac{1}{2}} &:& r\le \left((\nu-1)^2+1\right)^{\frac{1}{2}} \\
 \pm\left(1-\dfrac{\left(1-(\nu-1)\right)^2}{r^2}\right)^{1/2} &:& \left((\nu-1)^2+1\right)^{\frac{1}{2}} < r \le \nu\end{array}\right. \,.
\end{align}

In the expansion of double cones with the meridian curve $x(z)=1-|z|/\nu$ (see Fig.\ \ref{fig:critical_angles}),
more than two symmetrically equivalent intersections with spheres can exist for $r_c<r<1$ with $r_c=\sin(\arctan\nu)$ denoting the radius of the largest sphere completely inside the double cone.
In this region, the cosines of the critical pole distances read
\begin{align}
\xi_{1,2}=\cos\vartheta_{c_{1,2}}= \pm\dfrac{\nu}{(1+\nu^2)r}\left(1\pm \left(r^2\Big(1+\nu^2\right)-\nu^2\Big)^{\frac{1}{2}}\right).
\end{align}
For $1<r<\nu$ only two symmetrically equivalent intersections exist at the critical pole distances
\begin{align}
\xi=\cos\vartheta_{c}= \pm\dfrac{\nu}{(1+\nu^2)r}\left(1+\left(r^2\Big(1+\nu^2\right)-\nu^2\Big)^{\frac{1}{2}}\right).
\end{align}

\subsection{Expansion coefficients}

The angular-dependent function $\Psi(r,\vartheta,\varphi)$ can be expanded by using the expansion coefficients defined in Eq.\ \eqref{eq:flm}.
Since for rotationally symmetric solids of revolution, $\Psi(r,\vartheta)$ does not depend on the azimuthal angle $\varphi$,
the integration over $\varphi$ gives a nonzero contribution, namely a factor $2\pi$, only for $m=0$. For $m\neq 0$ the azimuthal integral vanishes so that only the coefficients $f_{l,0}$ are relevant. 

The solids of revolution considered here have an equatorial mirror plane, i.e., $\Psi(r,\vartheta)=\Psi(r,\pi-\vartheta)$.
Due to the parity
\begin{align}
P_l(x)&=(-1)^l P_l(-x)
\end{align}
of Legendre polynomials only coefficients with even $l$ differ from zero.

Thus, the relevant expansion coefficients for prolate particles are
\begin{align}
f_{2l}(r) &= 2\pi\int\limits_{-1}^1 \left(\dfrac{4l+1}{4\pi}\right)^\frac{1}{2} P_{2l}(x)\Psi(r,x)\;\rd x \nonumber \\
&=\left(\left(4l+1\right)\pi\right)^\frac{1}{2}\left(\int\limits_{-1}^{-1+\xi}P_{2l}(x)\;\rd x + \int\limits_{1-\xi}^1 P_{2l}(x)\;\rd x\right)\nonumber \\
&=\left(\left(4l+1\right)\pi\right)^\frac{1}{2}2\int\limits_{1-\xi}^1 P_{2l}(x)\,\rd x=2\left(\left(4l+1\right)\pi\right)^\frac{1}{2}\Xi_{2l}(\xi(r))\,,
\end{align}
where $x=\cos\vartheta$ and $\xi=\cos\vartheta_c$. The coefficients $\Xi(\xi(r))$ depend on the shape of the solids of revolution only via the cosine of the critical polar angles $\xi(r)=\cos\vartheta_c(r)$. 

\begin{figure}[tb]
\centerline{\includegraphics[width=0.45\textwidth]{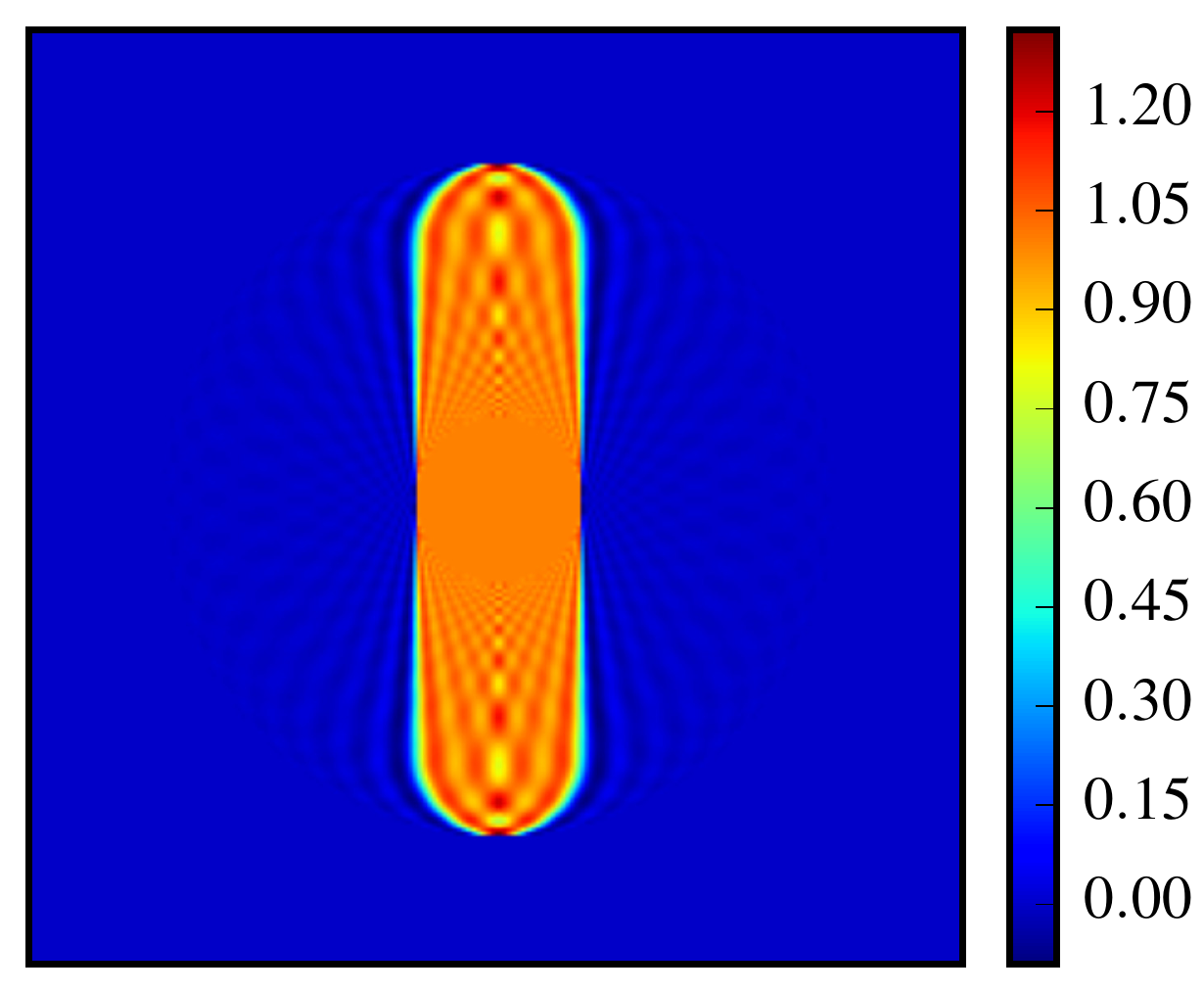}\qquad \quad \includegraphics[width=0.45\textwidth]{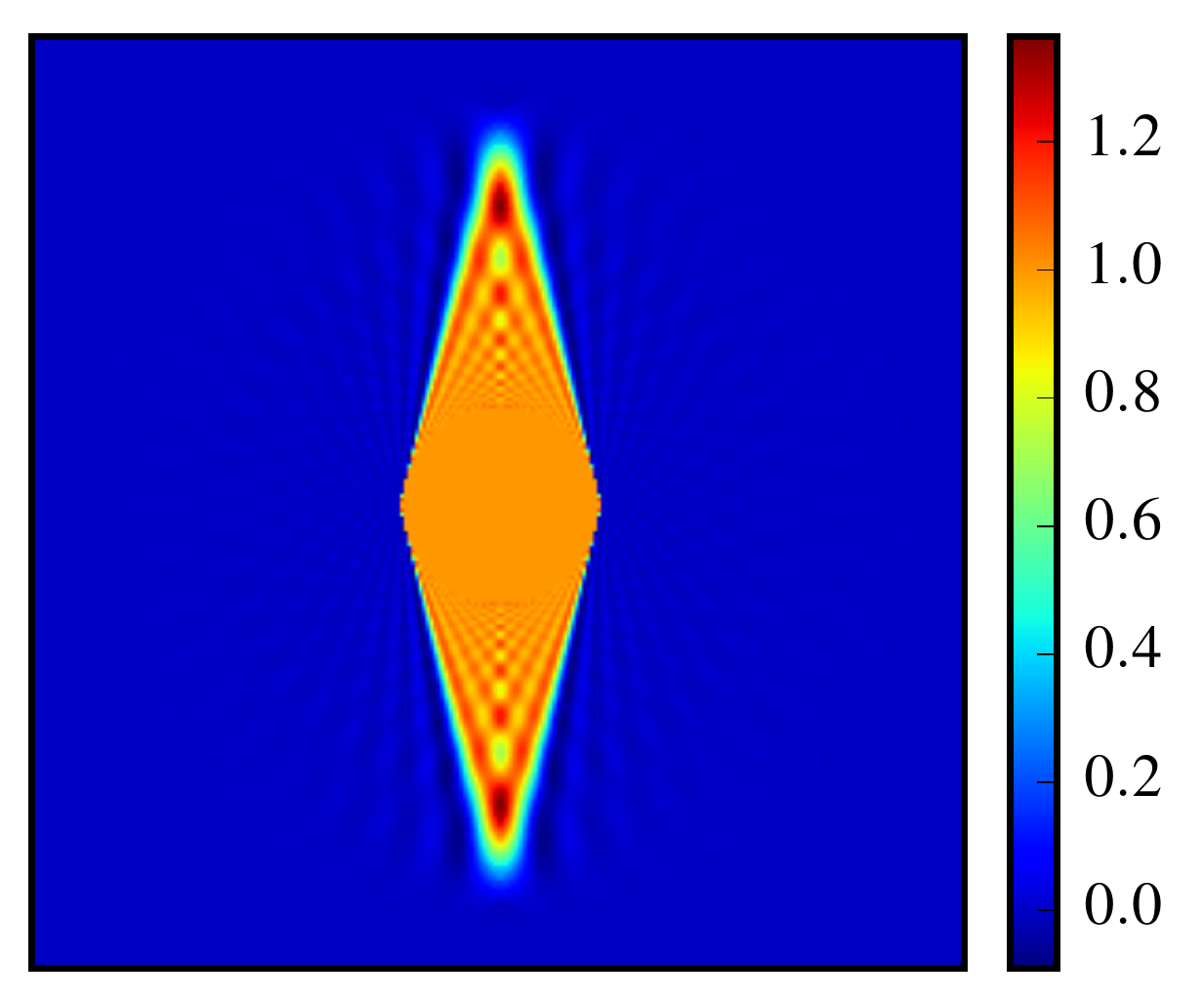}}
\caption{\label{fig:reconstruction}Reconstruction of a spherocylinder (lhs) and a double cone (rhs) with an aspect ratio of $\nu=5$ from the expansion in spherical harmonics up to the order $l=40$. In the scattering functions, the deviations from a uniform scattering length density are relevant for large scattering vectors probing the short length scales of correlations in the scattering length density.}
\end{figure}

For double cones, in the region $r_c<r<1$ additional intersections
centered at the equatorial plane occur, leading to additional contributions
\begin{align}
\Lambda_{2l}(\xi'(r)) &= 2\pi \left(\dfrac{4l+1}{4\pi}\right)^\frac{1}{2} \int\limits_{-\xi'}^{\xi'} P_{2l}(x)\,\rd x =2\left((4l+1)\pi\right)^\frac{1}{2} \int\limits_0^{\xi'} P_{2l}(x)\,\rd x 
\end{align}
to the expansion coefficients, which read in this case $f_{2l}(r)= 2((4l+1)\pi)^{1/2}(\Xi_{2l}(\xi(r))+\Lambda(\xi'(r))).$
As a test for the expansion in spherical harmonics, the meridional cross sections of the solids of revolution are reconstructed from the expansion coefficients $f_l(r)$. In Fig.\ \ref{fig:reconstruction}, the sum of the contributions up to the order $l=40$ is exemplarily displayed for spherocylinders and double cones.

\section{Scattering function}
\label{scattering_function}

The static scattering function of a particle with a homogeneous scattering length density can---normalized to the forward scattering---in terms of its form factor be written as
\begin{align}
P(\mathbf{Q})=P(Q,\boldsymbol{\Omega})&= \dfrac{1}{V^2}\left(\iiint\limits_V \Psi(\mathbf{r})\re^{\imath \mathbf{Q} \cdot \mathbf{r}}\,\rd^3\mathbf{r}\right) \left(\iiint\limits_V \Psi(\mathbf{r})\re^{-\imath \mathbf{Q}\cdot\mathbf{r}}\,\rd^3\mathbf{r}\right),
\end{align}

where $\mathbf{Q}$ denotes the scattering vector and $V$ the volume of the particle. The function $\Psi(\mathbf{r})$ is the
reduced scattering length density with the values $\Psi(\mathbf{r})=1$ inside and $\Psi(\mathbf{r})=0$ outside of the particle. The normalization to the squared particle volume relates the scattering intensity at the scattering vector $\mathbf{Q}$ to the forward scattering at $\mathbf{Q}=\mathbf{0}$.
\begin{widetext}
With $\Psi(\mathbf{r})=\Psi(r\boldsymbol{\Omega})$, where $\boldsymbol{\Omega}$ denotes the direction of $\mathbf{r}$, using the expansion in spherical harmonics the form factor reads
\begin{align}
P(Q,\vartheta,\varphi)&=\dfrac{1}{V^2}\left(\int\limits_0^\infty \oiint\limits_{4\pi}\sum\limits_{l}
f_{2l}(r)Y_{2l,0}^*(\vartheta,\varphi)\re^{\imath Q r\cos\vartheta}\,\rd \cos\vartheta\,\rd\varphi\,\rd r\right)\\\nonumber
&\qquad\times \left(\int\limits_0^\infty\oiint\limits_{4\pi}\sum\limits_{l'}
f_{2l}(r)Y_{2l,0}(\vartheta,\varphi)\re^{-\imath Q r\cos\vartheta}\,\rd \cos\vartheta\,\rd\varphi\,\rd r\right).
\end{align}
\end{widetext}
Here, the direction of $\mathbf{Q}$ is chosen to be the positive $z$-direction. Then the angle enclosed between $\mathbf{Q}$ and $\mathbf{r}$ is $\vartheta$. Under this assumption the expansion coefficients $f_{2l}(Q)$ in Fourier space can be obtained via Hankel transforms
\begin{align}
f_{2l}(Q)&=(\imath)^{2l}\int\limits_0^\infty 4\pi r^2 f_{2l}(r) j_{2l}(r)\,\rd r \,,
\end{align}
where $j_{2l}(r)$ denotes the spherical Bessel function of the order $2l$. Since these expressions do not depend
on angular coordinates anymore, the average over all orientations is directly obtained, and the result is independent of the choice of the laboratory coordinate system.

The scattered intensity at the scattering vector with the modulus $Q$ and the direction $\boldsymbol{\Omega}$ is
\begin{align}
P(Q,\boldsymbol{\Omega})&=\dfrac{1}{V^2}\left(\sum\limits_l f_{2l}(Q)Y_{2l,0}^*(\boldsymbol{\Omega})\right)\left(\sum\limits_{l'} f_{2l'}(Q)Y_{2l',0}(\boldsymbol{\Omega})\right).
\end{align}
Hence, for an ensemble of statistically aligned particles with the odf $p(\boldsymbol{\Omega})=1/(4\pi)$, the orientation-averaged form factor is given by
\begin{align}
\langle P(Q,\boldsymbol{\Omega})\rangle_{\boldsymbol{\Omega}} & = \dfrac{1}{V^2} \oiint\limits_{4\pi} \sum\limits_l\sum\limits_{l'}
f_{2l}(Q)f_{2l'}(Q) Y_{2l,0}^*(\boldsymbol{\Omega}) Y_{2l',0}^*(\boldsymbol{\Omega})\,\rd\boldsymbol{\Omega} = \dfrac{1}{V^2}\sum\limits_l f_{2l}^2(Q)\,,
\end{align}
since the odf eliminates due to the normalization to the forward scattering of an isotropic ensemble.

The form factors can be obtained via a different route by integration in cylinder coordinates.\cite{maerkert:2011} The semianalytic integration leads to form factors $P(Q,\vartheta)$ depending on the direction of the scattering vector $\hat{\mathbf{Q}}$ and the direction of the particle axis $\hat{\mathbf{r}}$  with $\cos\vartheta=\hat{\mathbf{Q}}\cdot \hat{\mathbf{r}}$.
The orientation average $P(Q)=\langle P(Q,\vartheta)\rangle_\vartheta$ can be calculated by numerical integration.

\begin{figure}[tb]
\centerline{\includegraphics[width=0.5\textwidth]{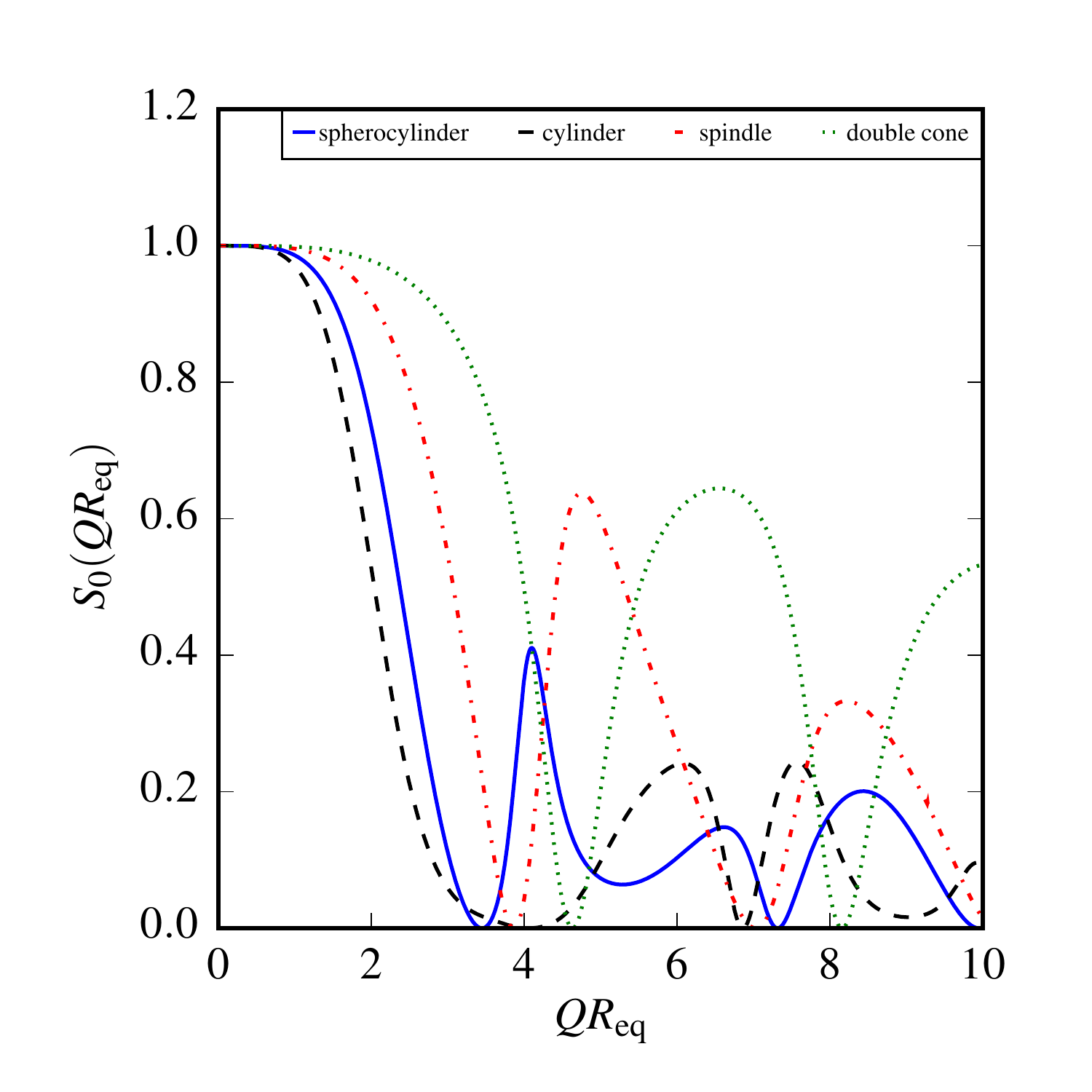}
\includegraphics[width=0.5\textwidth]{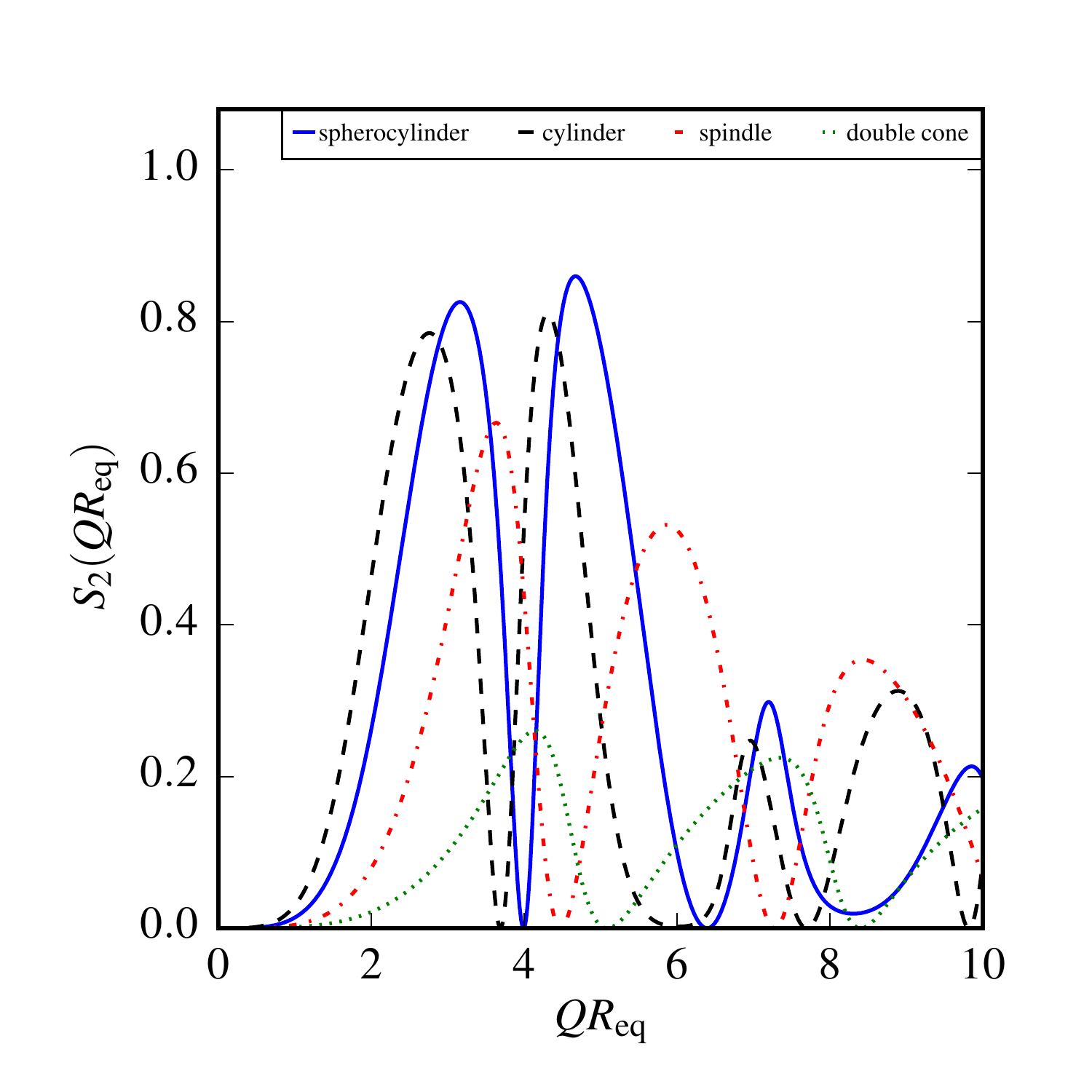}}
\centerline{\includegraphics[width=0.5\textwidth]{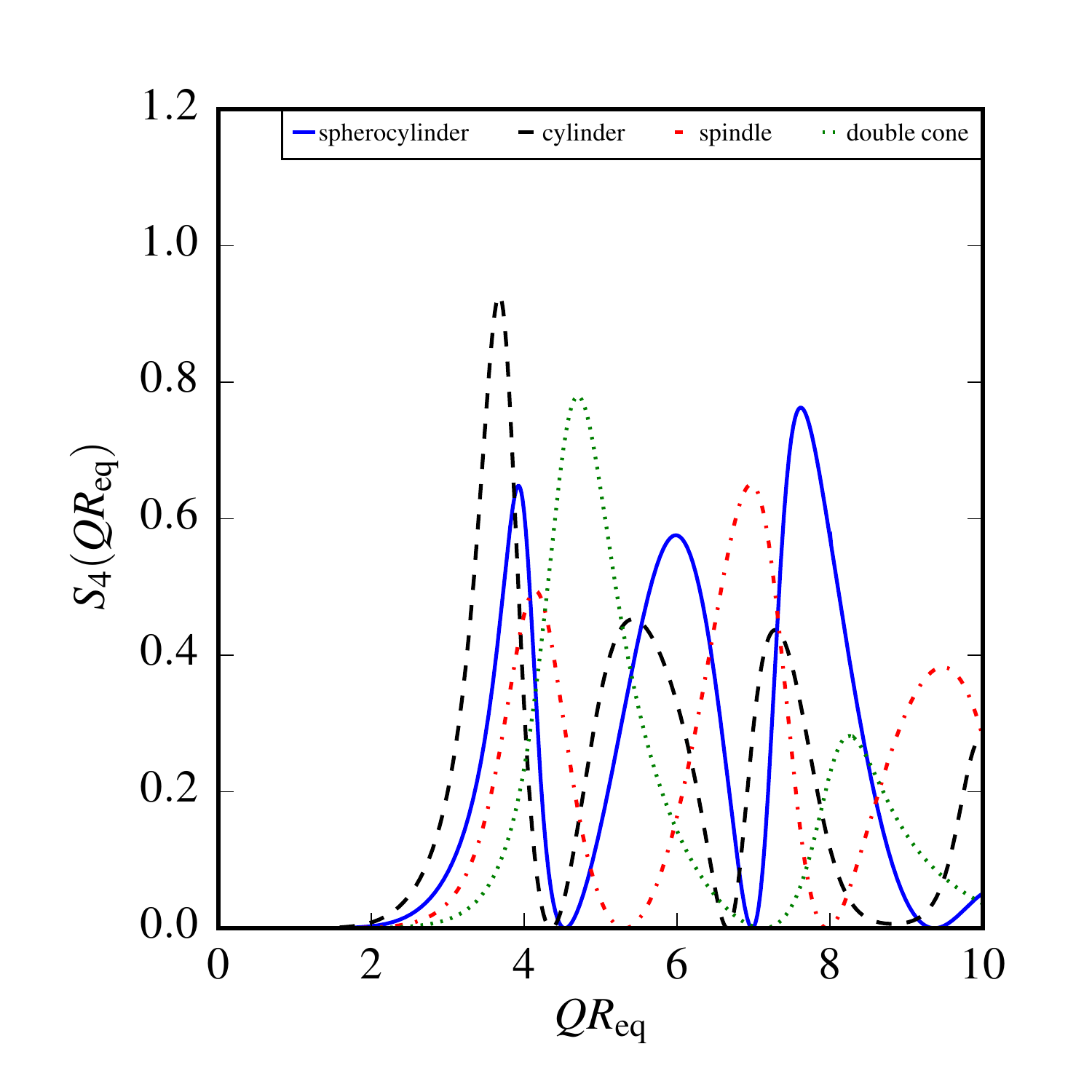}
\includegraphics[width=0.5\textwidth]{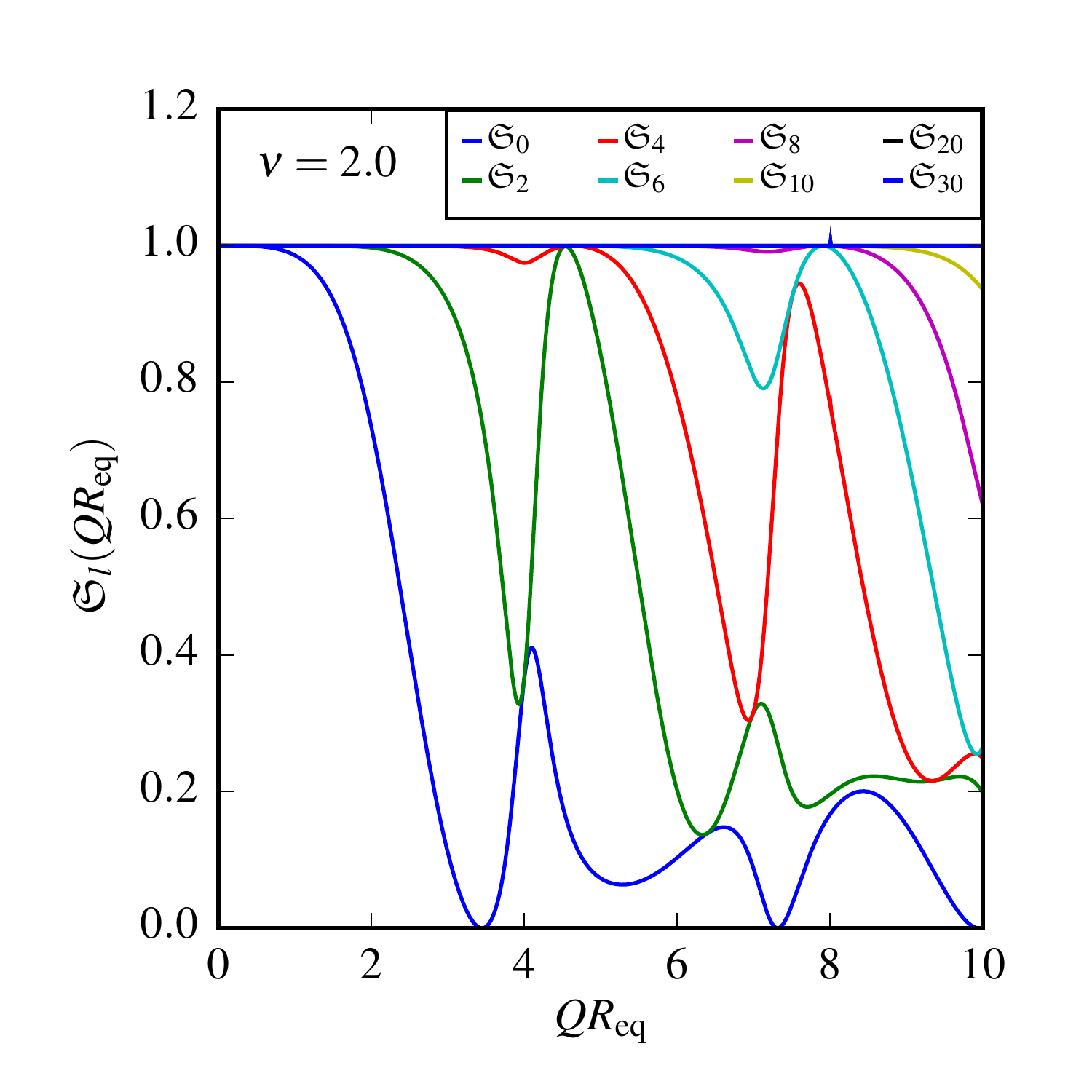}}
\caption{\label{fig:S_200}Expansion coefficients $S_{2l}(Q)$ for prolate solids of revolution with an aspect ratio of $\nu=2.0$ and a convergence test for spherocylinders with the aspect ratio $\nu=2.0$ (lower right corner). $QR_\mathrm{eq}$ is the reduced scattering vector with $R_{\rm eq}$ denoting the equatorial radius of the particles.}
\end{figure}

\begin{figure}[tb]
\centerline{\includegraphics[width=0.5\textwidth]{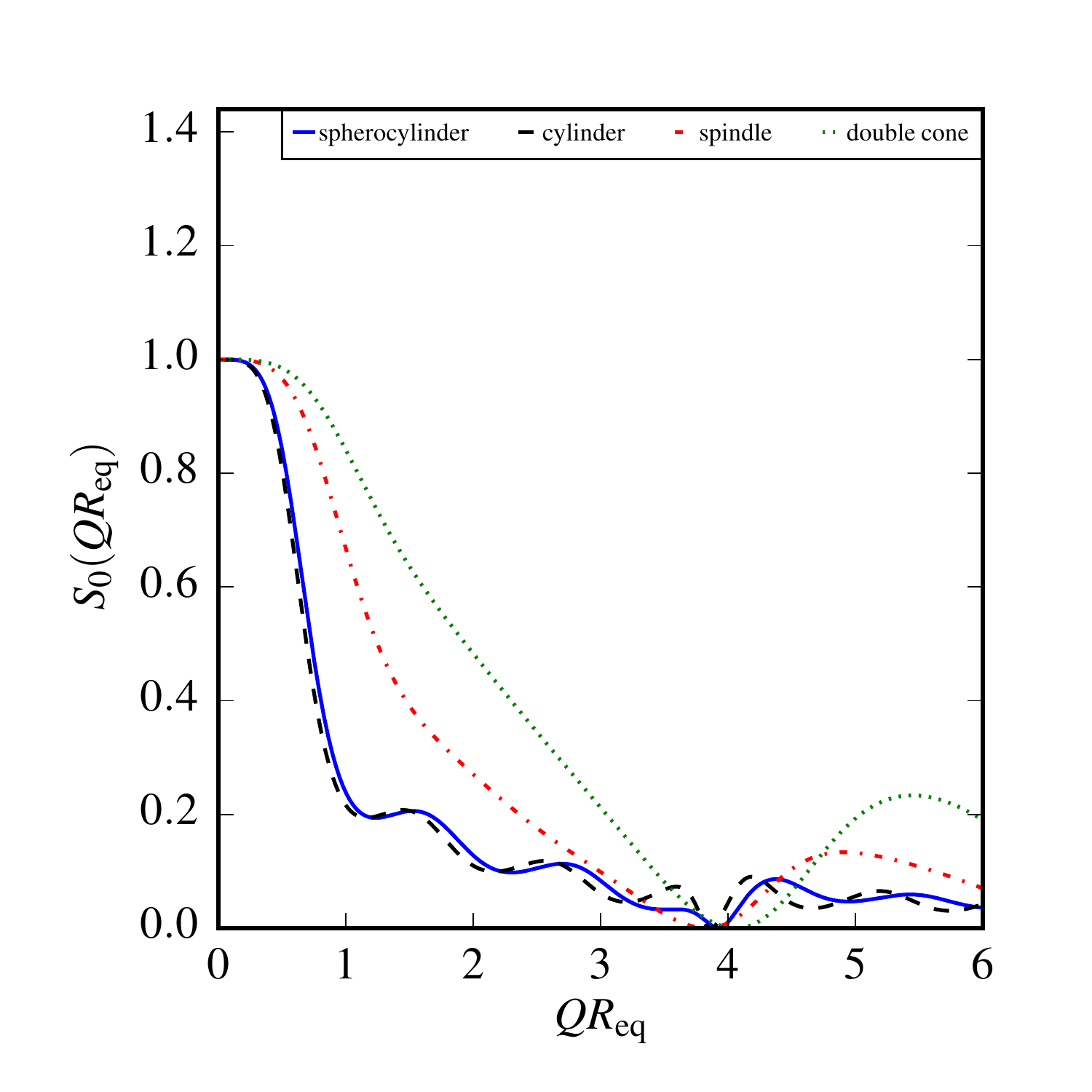}
\includegraphics[width=0.5\textwidth]{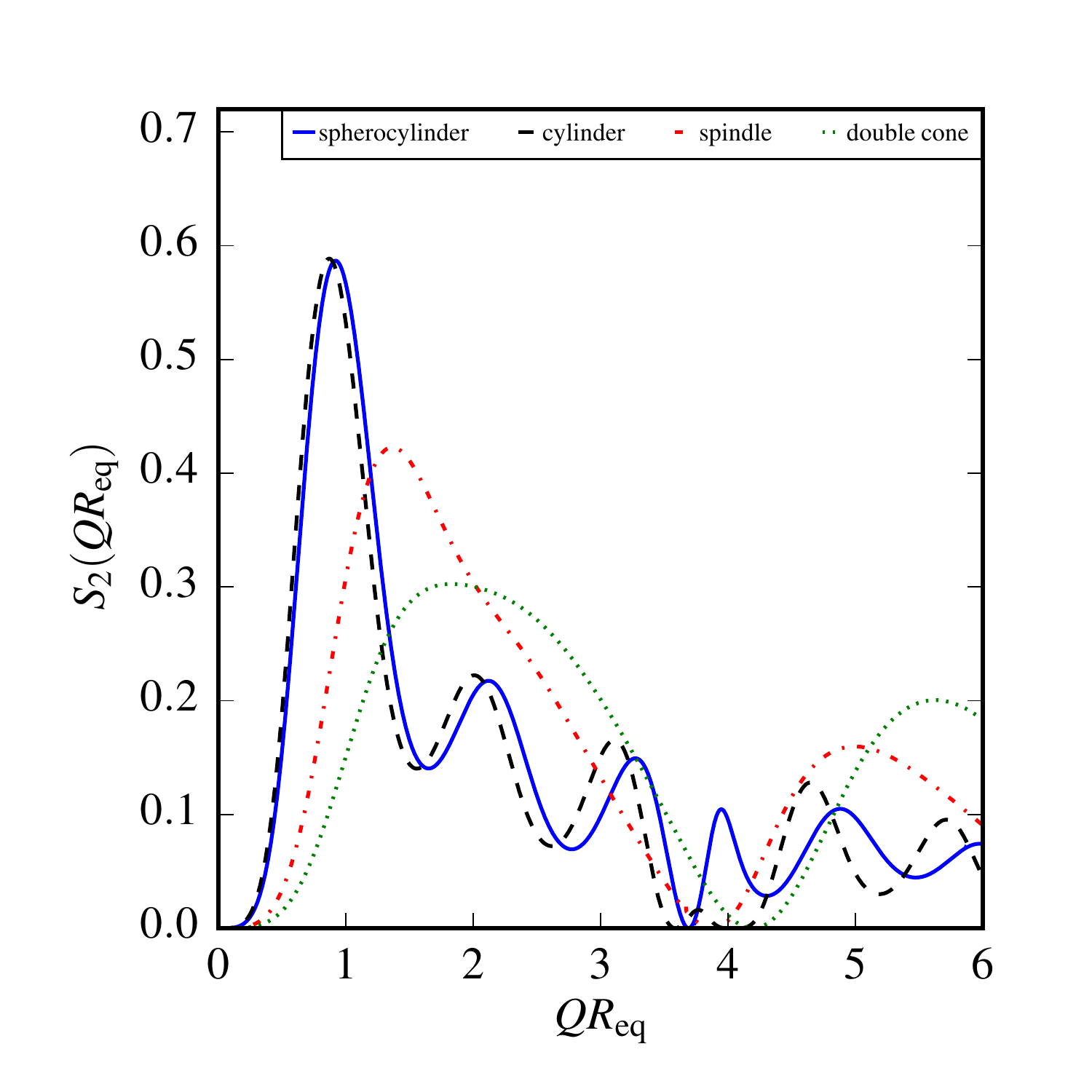}}
\centerline{\includegraphics[width=0.5\textwidth]{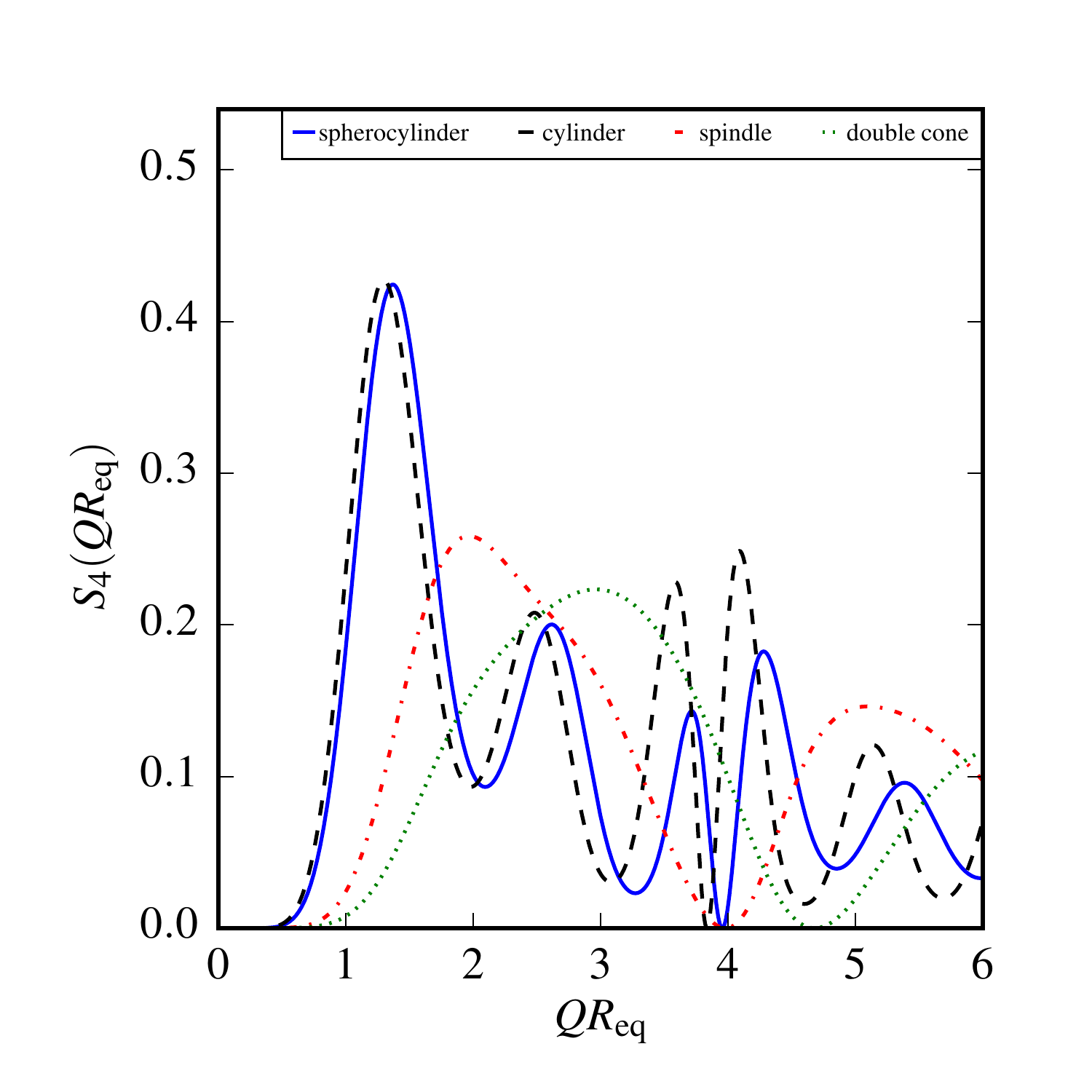}
\includegraphics[width=0.5\textwidth]{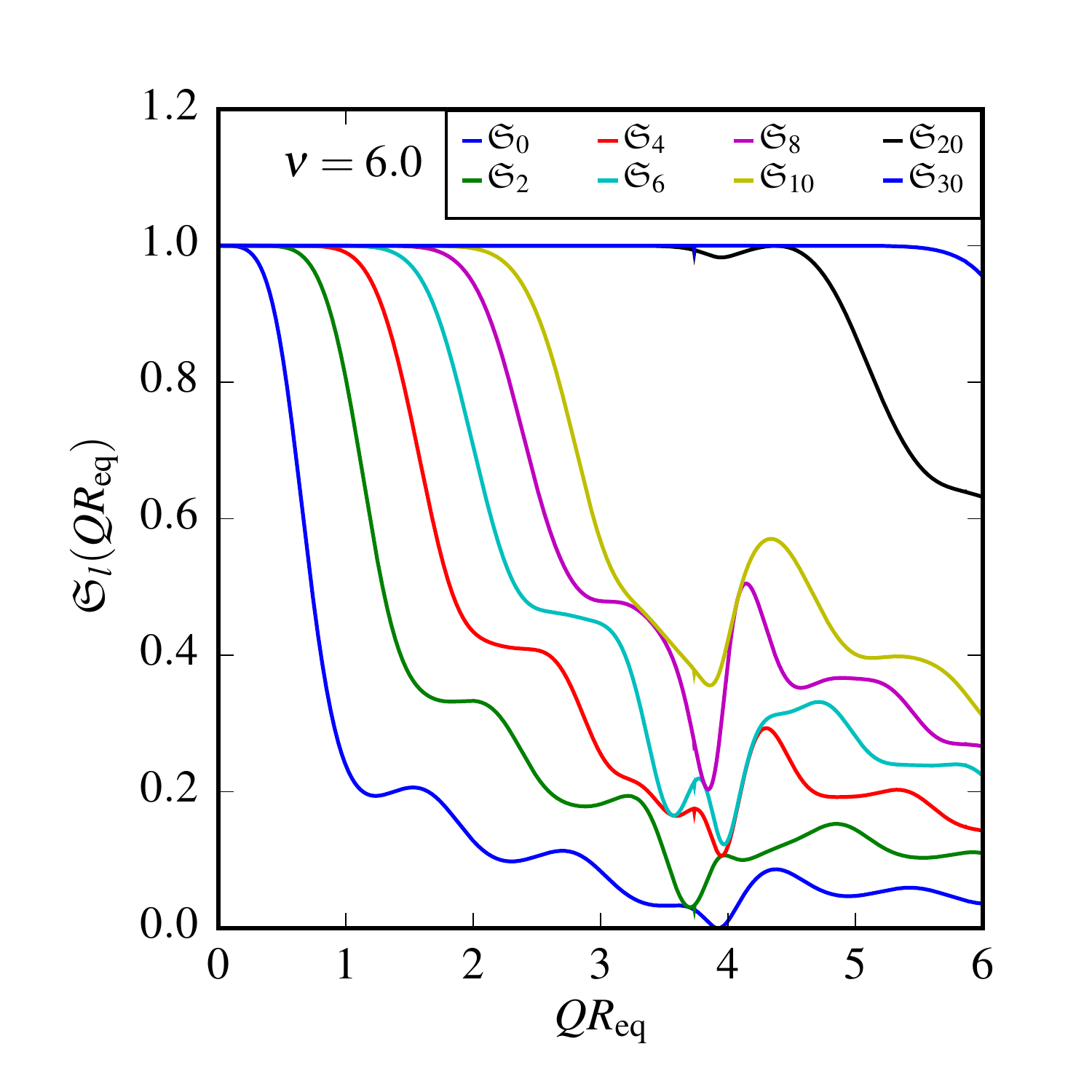}}
\caption{\label{fig:S_600}Expansion coefficients $S_{2l}(Q)$ for prolate solids of revolution with an aspect ratio of $\nu=6.0$ and a convergence test for spherocylinders with the aspect ratio $\nu=6.0$ (lower right corner). }
\end{figure}

Let us define reduced expansion coefficients 
\begin{align}
S_{2l}(Q)&=\dfrac{f_{2l}^2(Q)}{\sum\limits_l f_{2l}^2(Q)}=\dfrac{f_{2l}^2(Q)}{V^2\langle P(Q,\boldsymbol{\Omega})\rangle_{\boldsymbol{\Omega}}}\,.
\end{align}
Then, $S_0(Q)$ is formally the reduced monopole contribution to the scattering intensity, $S_2(Q)$ the quadrupole contribution, $S_4(Q)$ the hexadecapole contribution, and so on. 

If we further define
\begin{align}
\mathfrak{S}_l(Q)&= \sum\limits_{n=0}^l S_n(Q)
\end{align}
as the sum of the first $l$ expansion coefficients, the limit $\lim_{l\to \infty} \mathfrak{S}_l(Q) = 1$
is a test for the convergence of the expansion if $\langle P(Q,\boldsymbol{\Omega})\rangle_{\boldsymbol{\Omega}}$ is calculated via numerical integration in cylinder coordinates.

In Figs. \ref{fig:S_200} and \ref{fig:S_600}, the first expansion coefficients $S_0(Q)$, $S_2(Q)$, and $S_4(Q)$ are displayed for various particle shapes. We focus on cylinders, spherocylinders, spindles, and double cones since the convergence of the expansion of ellipsoids with large aspect ratios in terms of spherical harmonics is rather unrewarding. The plots in the lower right corners of Figs. \ref{fig:S_200} and \ref{fig:S_600} illustrate the convergence of the expansion exemplarily for spherocylinders by showing different $\mathfrak{S}_l(Q)$. For highly anisotropic objects, the limitation of the $Q$-range is visible even up to the order $l=30$. Expansion coefficients of higher order lead to numerical instabilities in the numerical Hankel transform with double precision arithmetics.

\begin{figure}[tb]
\centerline{\includegraphics[width=0.5\textwidth]{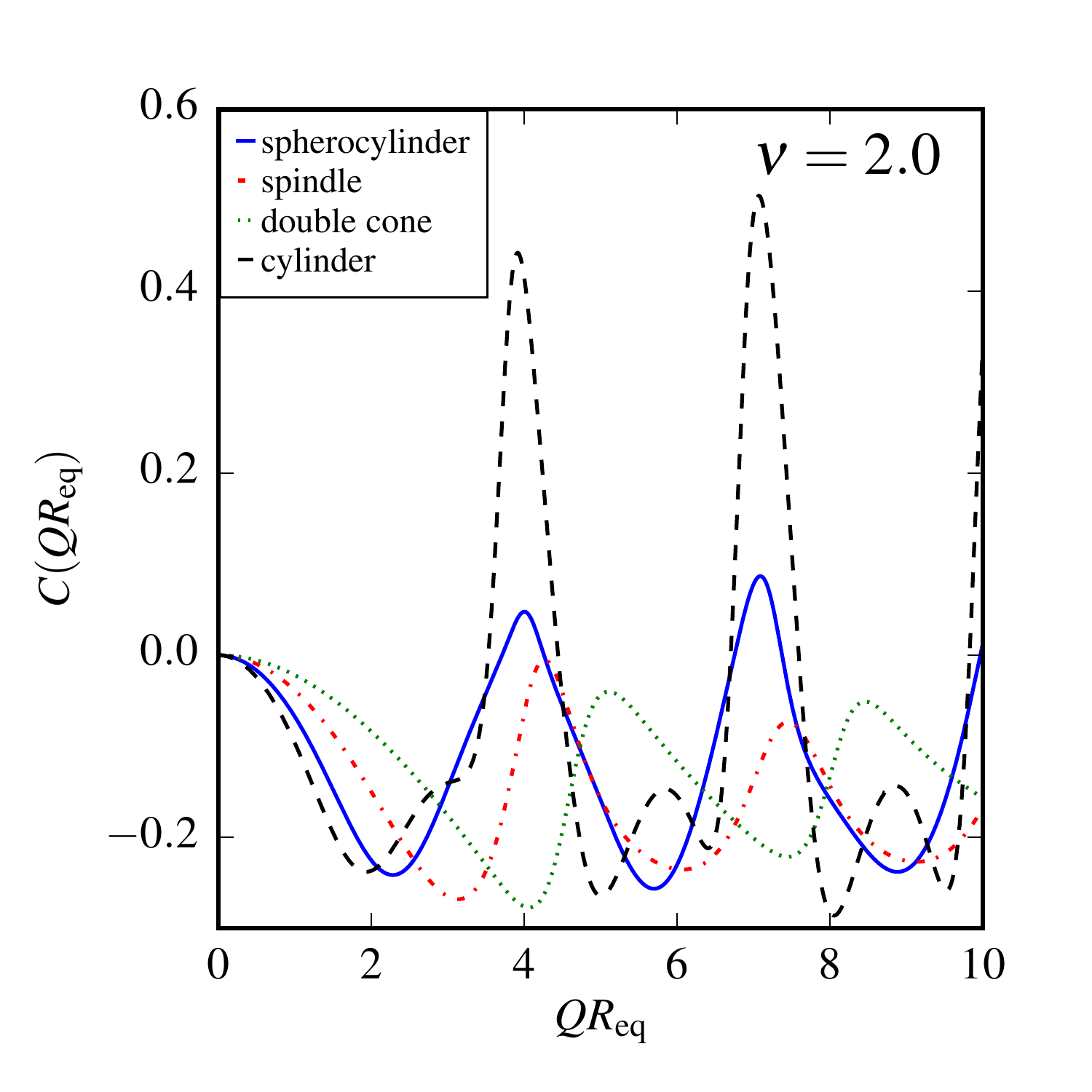}\includegraphics[width=0.5\textwidth]{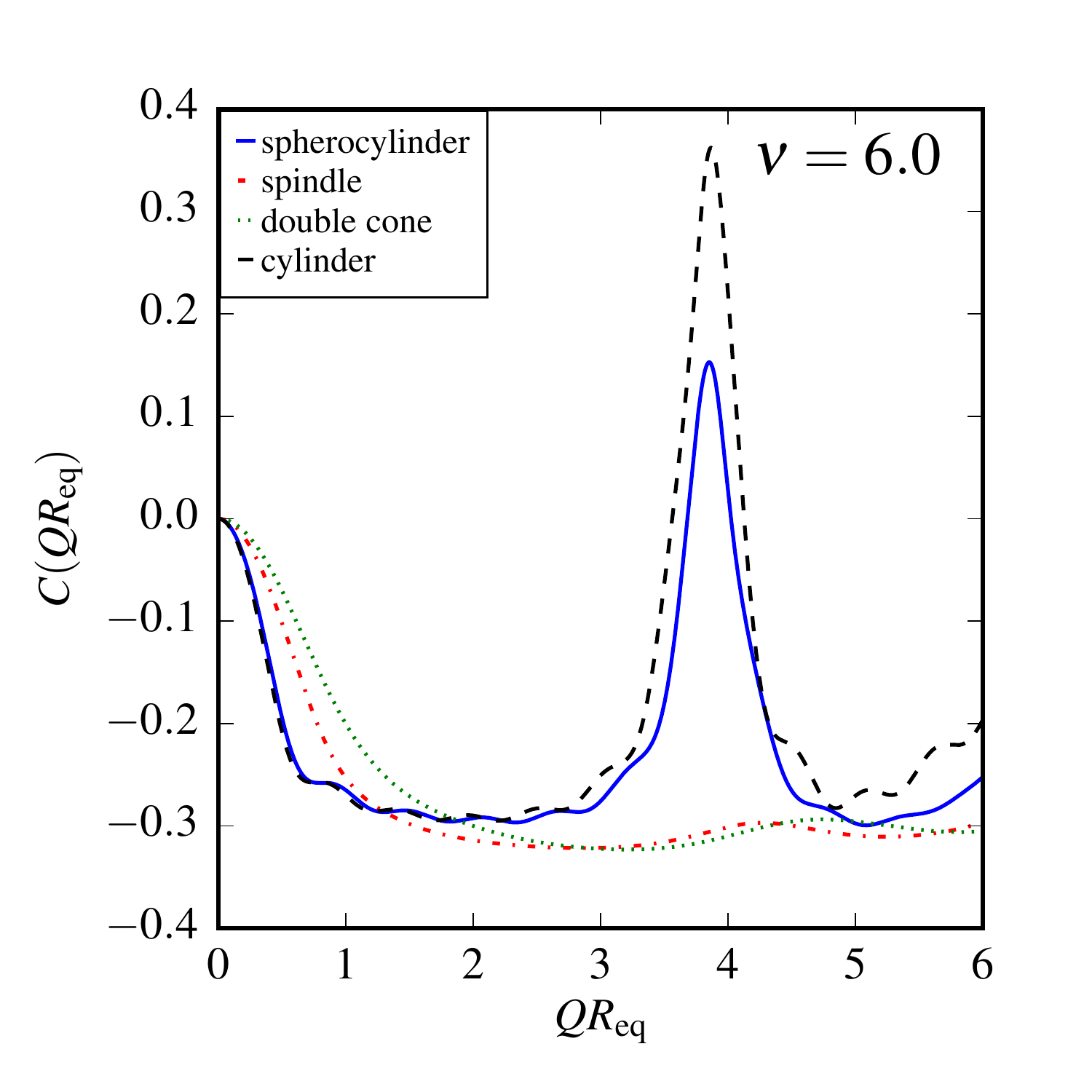}}
\caption{\label{fig:coupling_functions} Coupling functions $C(Q)$ for cylinders, spherocylinders, spindles, and double cones with aspect ratios $\nu=2.0$ (lhs) and $\nu=6.0$ (rhs). }
\end{figure}

The coupling function $C(Q)$ defined in Eq. \eqref{eq:coupling_function}  also depends on the expansion coefficients $f_{2l}(Q)$ weighted by Gaunt coefficients. Coupling functions for differently shaped objects are displayed in Fig.\ \ref{fig:coupling_functions} for the aspect ratios $\nu=2$ and $\nu=6$. The increasing similarity of $C(Q)$ for cylinders and spherocylinders with increasing aspect ratio is reasonable since the differences in the shape vanish in this case.
Only at large scattering vectors probing short intraparticle correlation lengths, significant differences exist.

Combining Eqs. \eqref{eq:autocorrelation_rot}, \eqref{eq:autocorrelation_daverage}, \eqref{eq:autocorrelation_danisotropic}, and \eqref{eq:prod} leads to the expression
\begin{align}
\label{eq:scat}
g_1^{\rm VH}(Q,t) &= \exp\left(-Q^2\,\langle D_{\rm tr} \rangle t\right)  \exp\left(-C(Q) Q^2 \Delta D_{\rm tr}\,t\right) \sum\limits_l S_l(Q)\exp\left(-l(l+1) D_{\rm rot}t\right) 
\end{align}
for the total depolarized field autocorrelation function since, due to the rotational symmetry of the particles only expansion coefficients $S_l(Q)=f_{l,0}^2(Q)/(V^2\langle P(Q,\boldsymbol{\Omega})\rangle_{\boldsymbol{\Omega}})$  are nonzero.

Experimental access to the short-time behavior of the intermediate scattering function $g_1^{\rm VH}(Q,t)$
is enabled by its first cumulants $\Gamma(Q)$, i.e., the negative initial slope with respect to $t$,

\begin{align}
\label{eq:cumu}
\Gamma(Q) &= -\lim\limits_{t\to 0} \dfrac{\partial g_1^{\rm VH}(Q,t)}{\partial t} = -\lim\limits_{t\to 0}\dfrac{\partial\ln g_1^{\rm VH}(Q,t)}{\partial t}\\ & = Q^2\langle D\rangle_{\rm tr} +Q^2\Delta D_{\rm tr} C(Q) + \sum\limits_l l(l+1) D_{\rm rot}S_l(Q)\,.
\nonumber
\end{align}
While in a polarized scattering experiment in VV-geometry the first cumulants are proportional to $Q^2$ according to the Landau-Placzek relation,\cite{landau:1934} in a depolarized scattering geometry two additional contributions exist. On the one hand, due to the anisotropy of the translational diffusion, an additional contribution proportional to $Q^2$, which is however modulated with the coupling function $C(Q)$, occurs. On the other hand, the first cumulants are influenced by the rotational diffusion.

\section{\label{coefficients}Calculation of the diffusion coefficients}

The intermediate scattering function $g_1^{\rm VH}(Q,t)$ [see Eq.\ \eqref{eq:scat}] as well as its first cumulants $\Gamma(Q)$ [see Eq.\ \eqref{eq:cumu}] depend on the translational and rotational diffusion coefficients of the considered particles. We calculate these coefficients for various particle shapes based on a bead model using the software HYDRO++.\cite{delaTorre2007,delaTorre1994,Carrasco99} 
This software has originally been designed for the prediction of hydrodynamic properties of macromolecules,\cite{delaTorre1994} but it can also be used for colloidal particles \cite{Dhont_book} modeled by a large number of beads.\cite{KaiserPRL2014}
To calculate the diffusion coefficients of such particles, first the bead positions have to be determined. In order to achieve a high accuracy, it is important that the bead model closely represents the specific particle shape. As the number of beads that can be used for the calculations with the software HYDRO++ is limited, it is most efficient to place the beads only on the surface of the particle to be modeled.  This method is referred to as \textit{shell model} and was originally introduced by Bloomfield.\cite{Bloomfield1967a,Bloomfield1967b} It enables a much better resolution of the particle shapes than filling models with a full-volume bead representation as beads with a significantly smaller radius can be used. Moreover, in many situations shell models even provide more accurate results than filling models with the same bead size.\cite{CarrascoBiophys1999} Best results are obtained if the beads are placed on the surface without larger voids so that they ideally touch each other. However, it is important that the beads do not overlap because the software may provide wrong results in that case. If bead representations with overlapping spheres are required, the program HYDROSUB can be used to perform the hydrodynamic calculations.\cite{delaTorreC2002,Kraft:13} This software also allows for the implementation of ellipsoidal and cylindrical subunits.\cite{delaTorreC2002} In order to include the effect of a wall or a substrate, it is in principle possible to modify the calculations by using the Stokeslet close to a no-slip boundary.\cite{Blake71} This is in particular relevant to experimental situations with a quasi-two-dimensional setup.\cite{Kuemmel:13,tenHagenKWTLB2014}

As we are dealing with rotationally symmetric particles, we apply the following efficient method to calculate the bead positions that are required as an input for HYDRO++.\footnote{The software HYDRO++ can also be used to calculate the diffusion coefficients of biaxial particles. In that case, it becomes a bit more complicated to calculate the bead positions for the bead models but this does not nameable affect the computation time.} The beads are placed on several rings centered on the $z$-axis, which coincides with the symmetry axis of the particles. These rings have a distance of $2r$ each, where $r$ is the radius of a single bead. We always use the same value of $r$ for all beads representing a particle. The radius $R(z)$ of the rings determining the bead positions is given by the meridian curves presented in Sec.\ \ref{expansion} and in Appendix A, respectively. 
For cylinders, it is necessary to add beads on the top and bottom areas, as otherwise beads are only placed on the lateral surface area of the cylinder. Therefore, additional rings with radii equidistantly distributed between zero and the equatorial radius $R_\mathrm{eq}$ are considered in this case.

\begin{figure}[tb]
\centering
\includegraphics[width=\columnwidth]{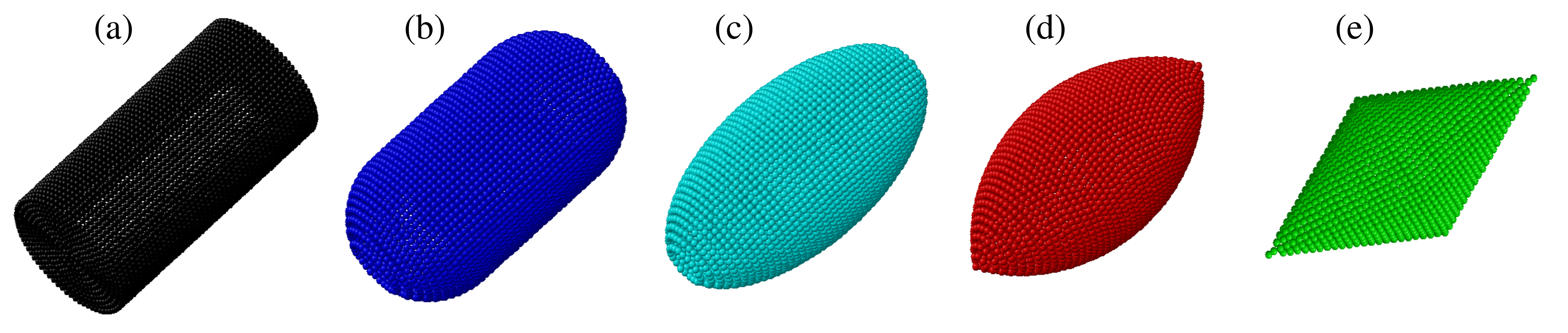}
\caption{\label{fig:bead_prolatec}Bead model representations of the considered prolate particle shapes with aspect ratio $\nu=2$: (a) cylinder, (b) spherocylinder, (c) ellipsoid, (d) spindle, and (e) double cone.}
\end{figure}

\begin{figure}[tb]
\centerline{\includegraphics[width=0.5\textwidth]{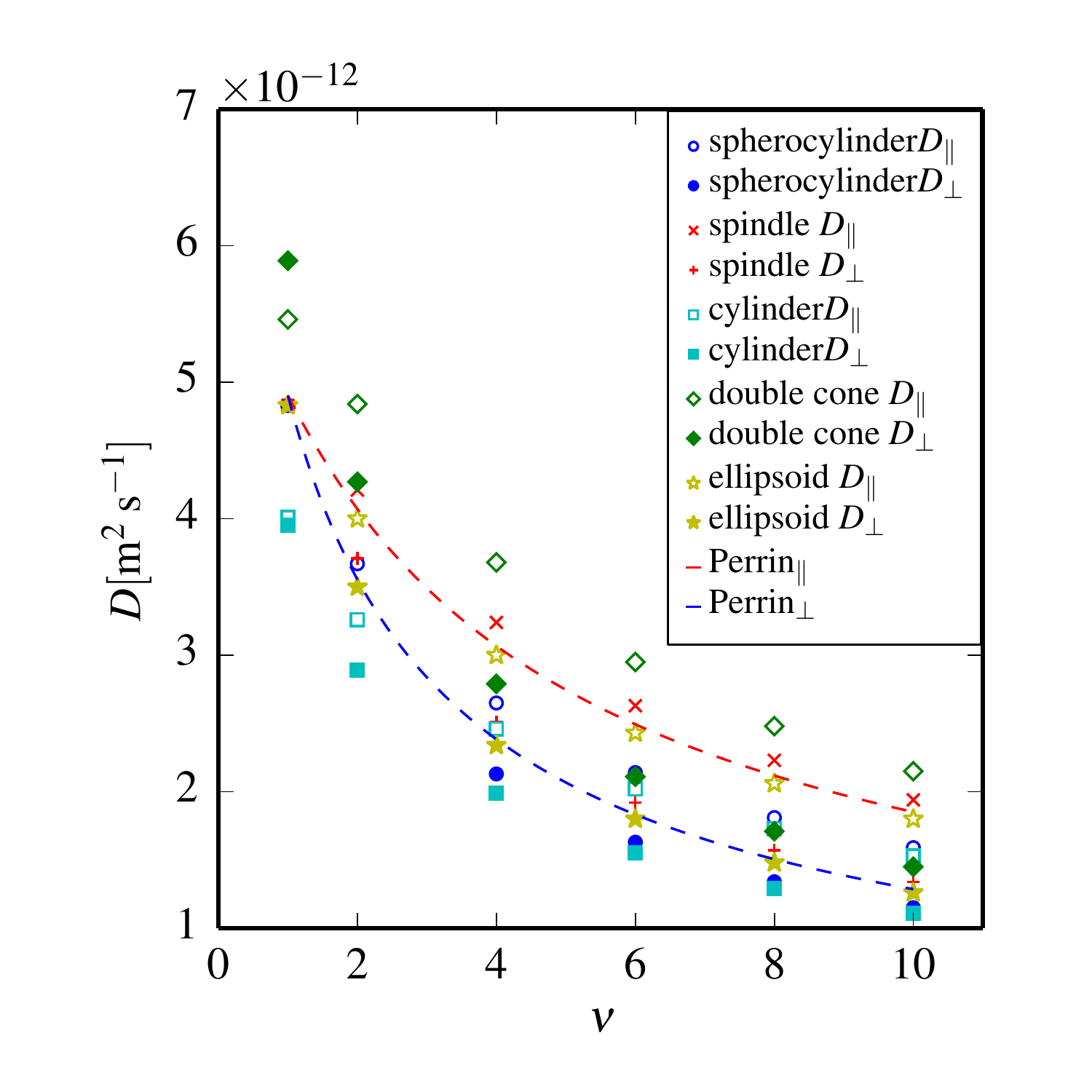}
\includegraphics[width=0.5\textwidth]{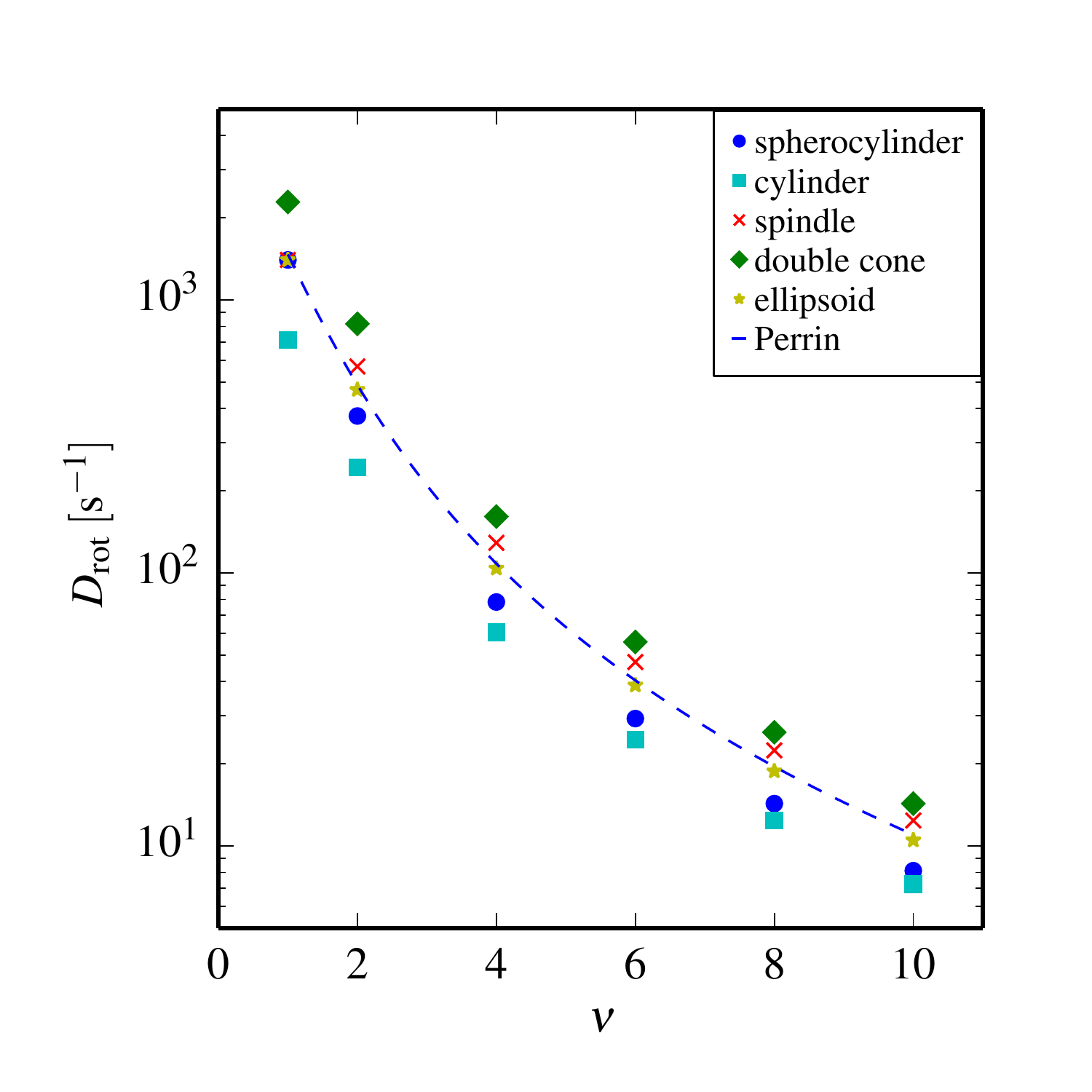}}
\caption{\label{fig:D_tensor}Translational (lhs) and rotational (rhs) diffusion coefficients for cylinders, spherocylinders, ellipsoids, spindles, and double cones with \mbox{$R_{\rm eq}=50$ nm} in dependence on the aspect ratio $\nu$. The dashed lines represent the analytically derived diffusion coefficients for ellipsoids.\cite{perrin:1934}}
\end{figure}

Once the radii of the rings are known, the explicit coordinates of the beads can be calculated. The maximum number of beads that can be placed on a ring with radius $R$ is determined by means of the formula $s=2R\sin(\alpha/2)$ for the chord length $s$. Here, $\alpha$ is the corresponding central angle. Using the condition $s=2r$ one obtains that the minimal ring radius 
\begin{equation}   
R_\mathrm{min}(N)=\frac{r}{\sin(\pi/N)}
\label{eq:beadnumber}
\end{equation}
is required in order to place $N$ beads of radius $r$ on the ring. Based on Eq.~(\ref{eq:beadnumber}), for each value of $R(z)$ the maximum number of beads can easily be determined. Using polar coordinates, the obtained number of beads are positioned equidistantly on the ring. The resulting bead model representations for the various considered particle shapes are visualized in Fig.\ \ref{fig:bead_prolatec} ,while the obtained values for the translational and rotational diffusion coefficients are displayed in Fig.\ \ref{fig:D_tensor}.

\section{\label{cumulants}Discussion of the first cumulants}

\begin{figure}[tb]
\centerline{\includegraphics[width=0.5\textwidth]{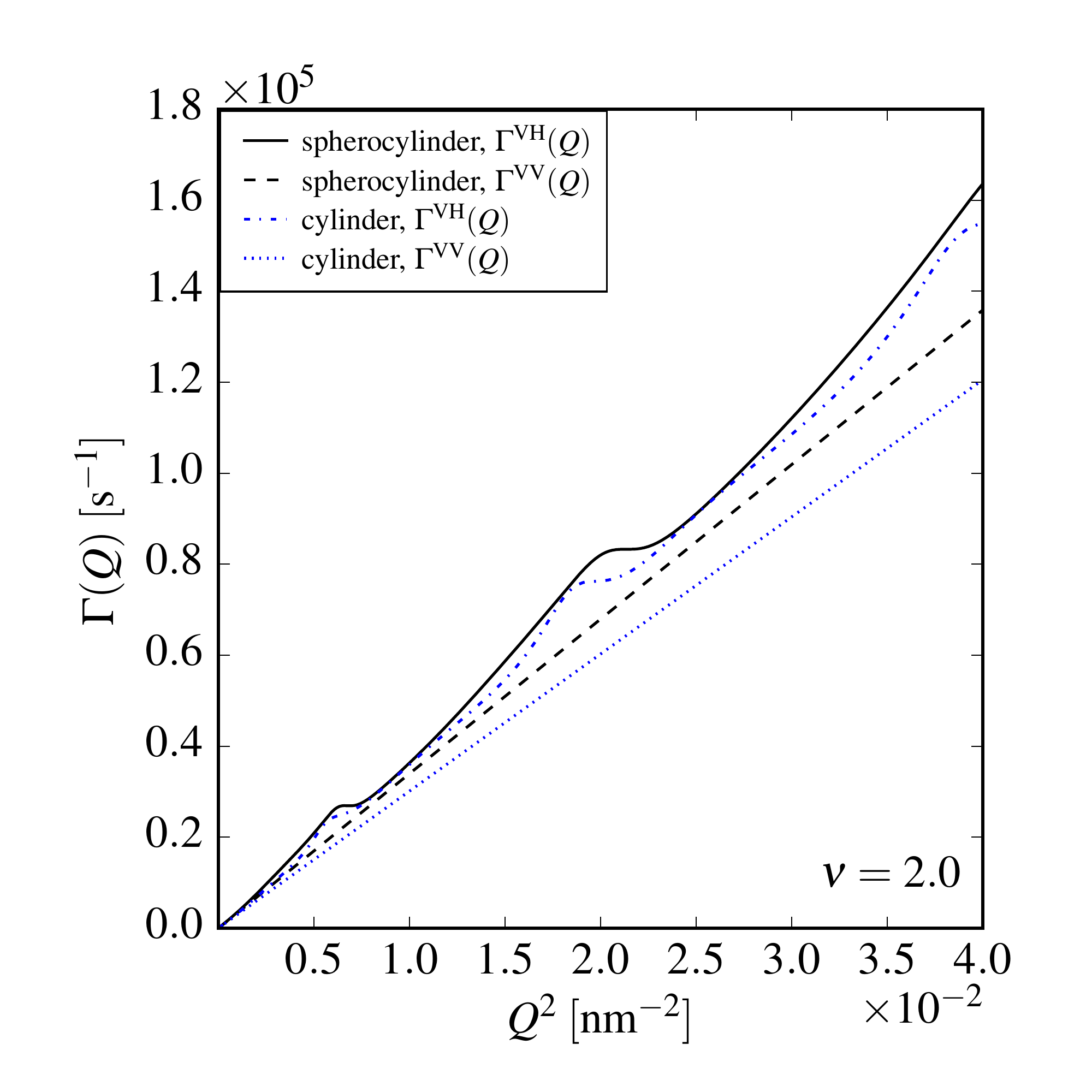}
\includegraphics[width=0.5\textwidth]{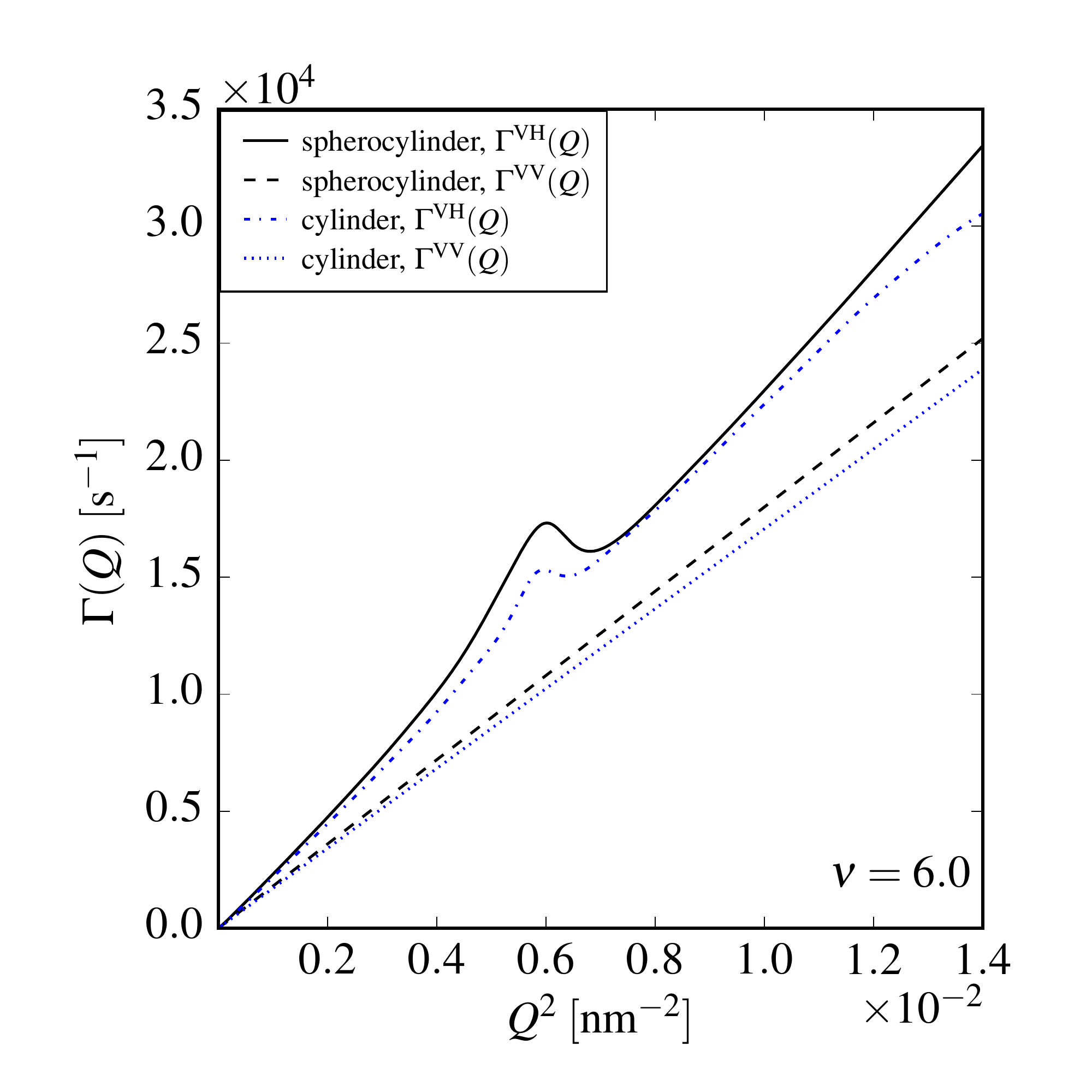}}
\centerline{\includegraphics[width=0.5\textwidth]{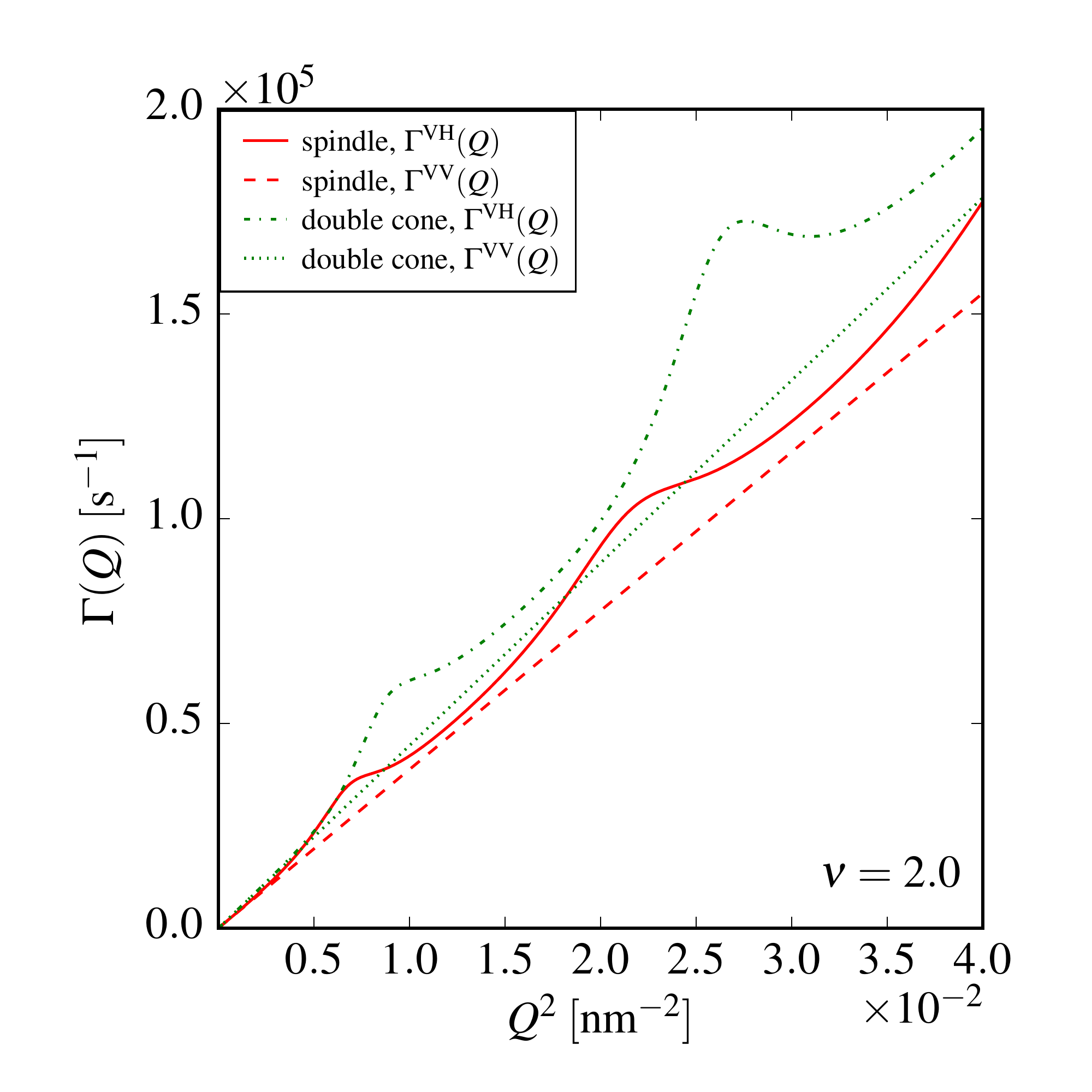}
\includegraphics[width=0.5\textwidth]{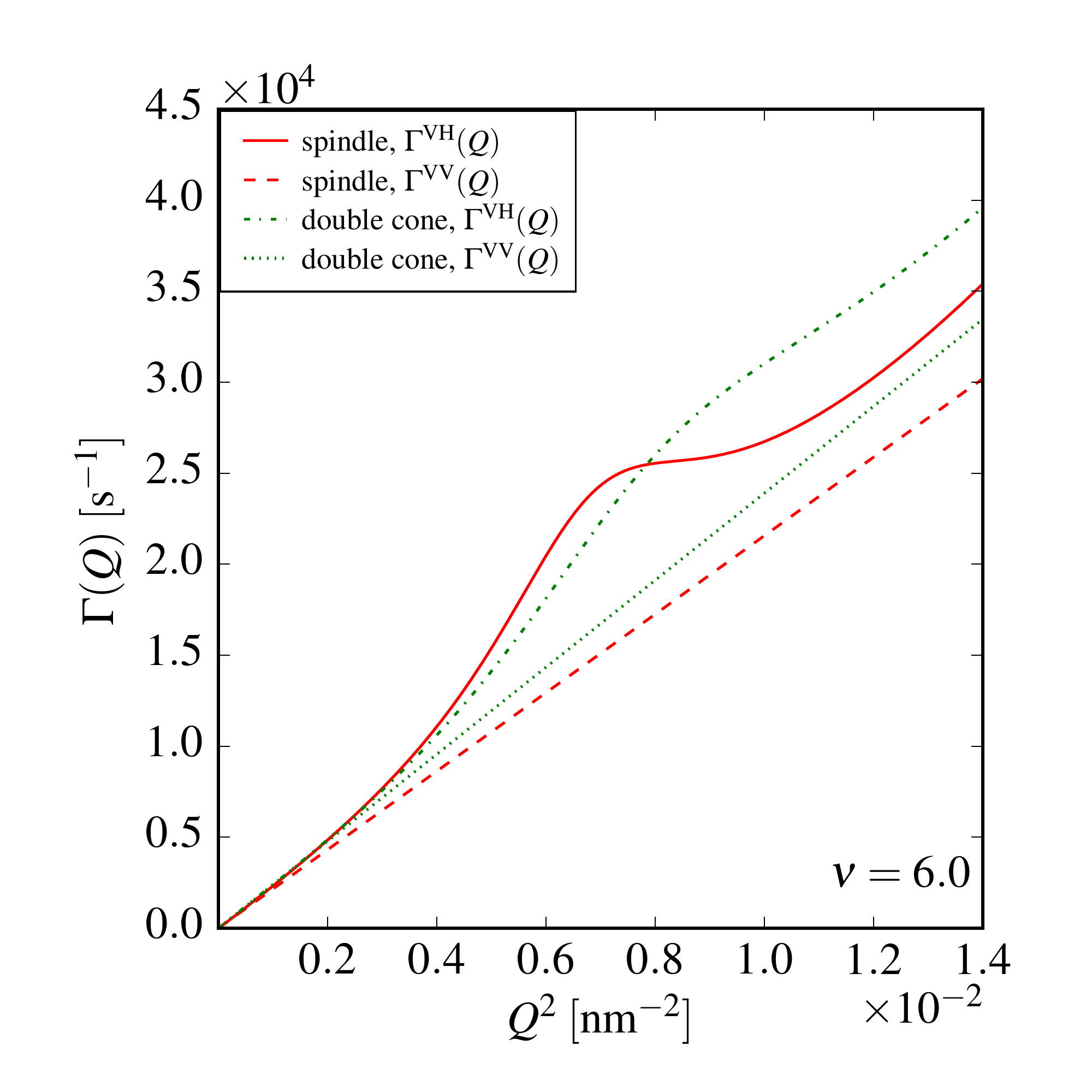}}
\caption{\label{fig:cumulants}First cumulants for differently shaped particles with aspect ratios of $\nu=2$ (lhs) and $\nu=6$ (rhs). The first cumulants of depolarized scattering $\Gamma^{\rm VH}(Q)$ are compared to those resulting from polarized scattering $\Gamma^{\rm VV}(Q)$.}
\end{figure}

Using the diffusion coefficients shown in Fig.\ \ref{fig:D_tensor}, the cumulants for polarized and depolarized scattering experiments can be quantitatively analyzed. They are displayed in Fig.\ \ref{fig:cumulants} for particles with an aspect ratio of $\nu=2$ and $\nu=6$, respectively. For objects with typical lengths comparable or larger than the wavelength, also the rotational contribution depends on the scattering vector due to the formal multipole expansion of the scattering intensity.

The first cumulants as the initial slope of the field autocorrelation function provide the dynamic quantity which can be accessed with the highest accuracy from experimental data within the regime of the lowest signal to noise ratio. Since the coupling function is calculated as a short time expansion, in addition, the initial slope of the  autocorrelation function is well described by the theoretical approach discussed in this paper.

\begin{figure}[tb]
\centerline{\includegraphics[width=0.5\textwidth]{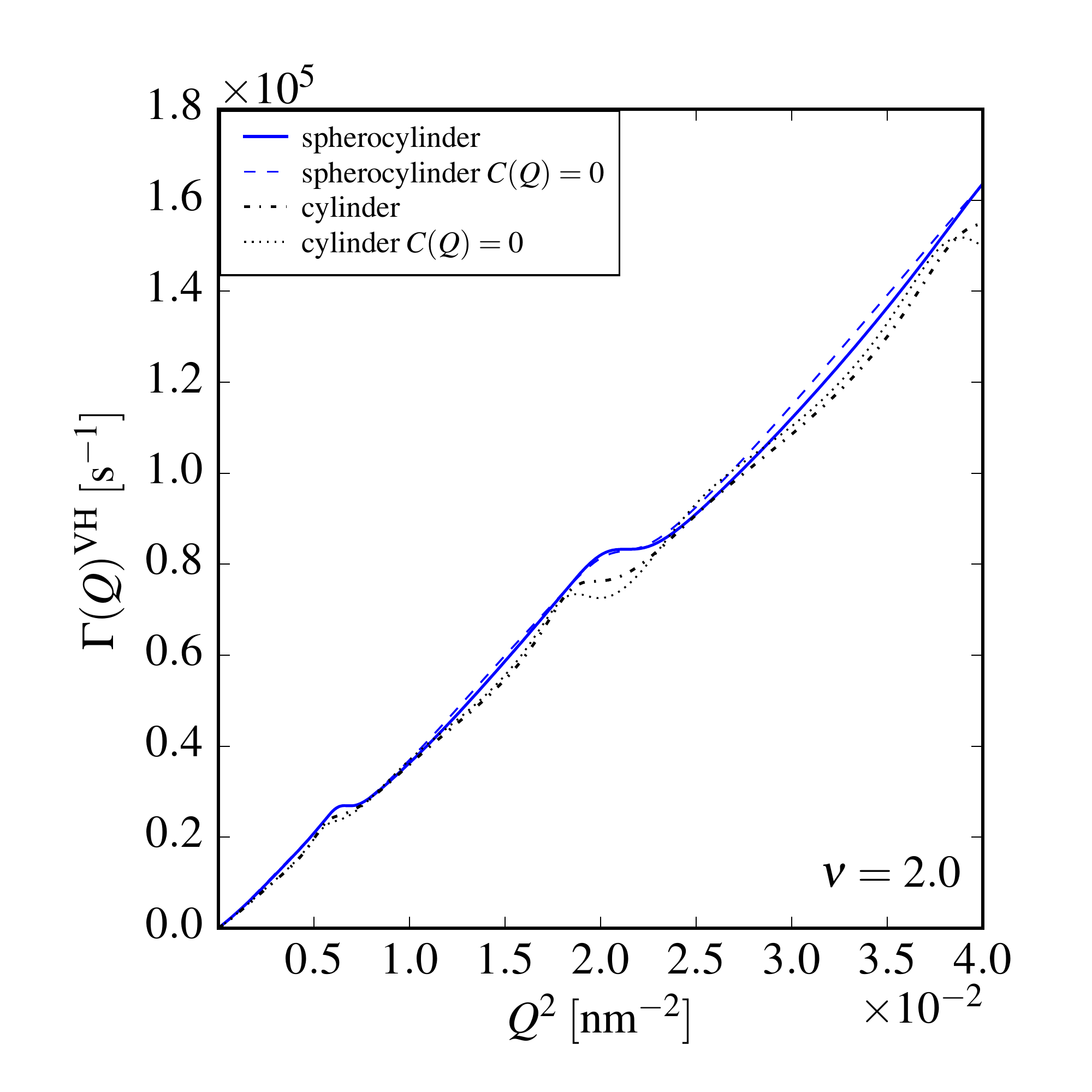}
\includegraphics[width=0.5\textwidth]{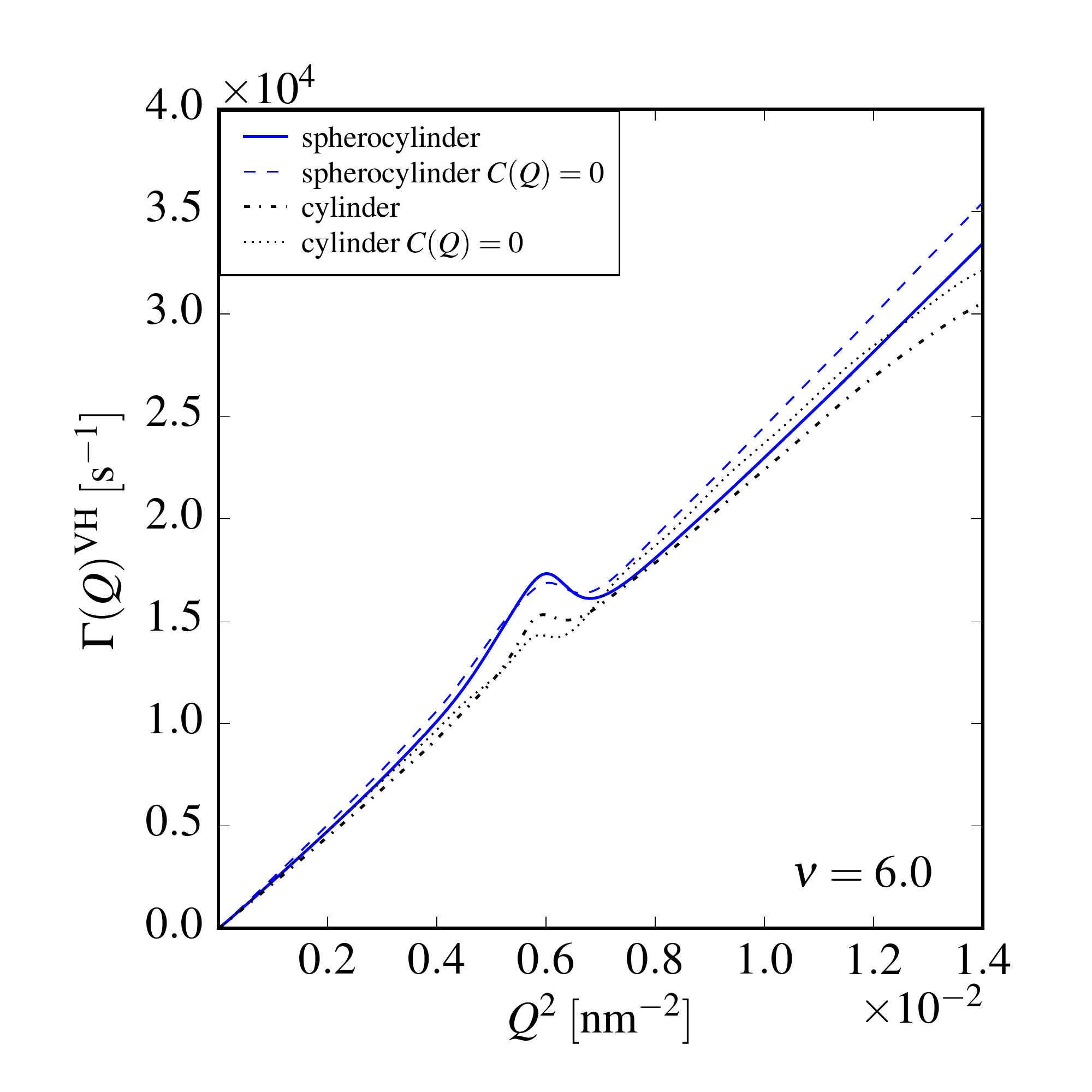}}
\centerline{\includegraphics[width=0.5\textwidth]{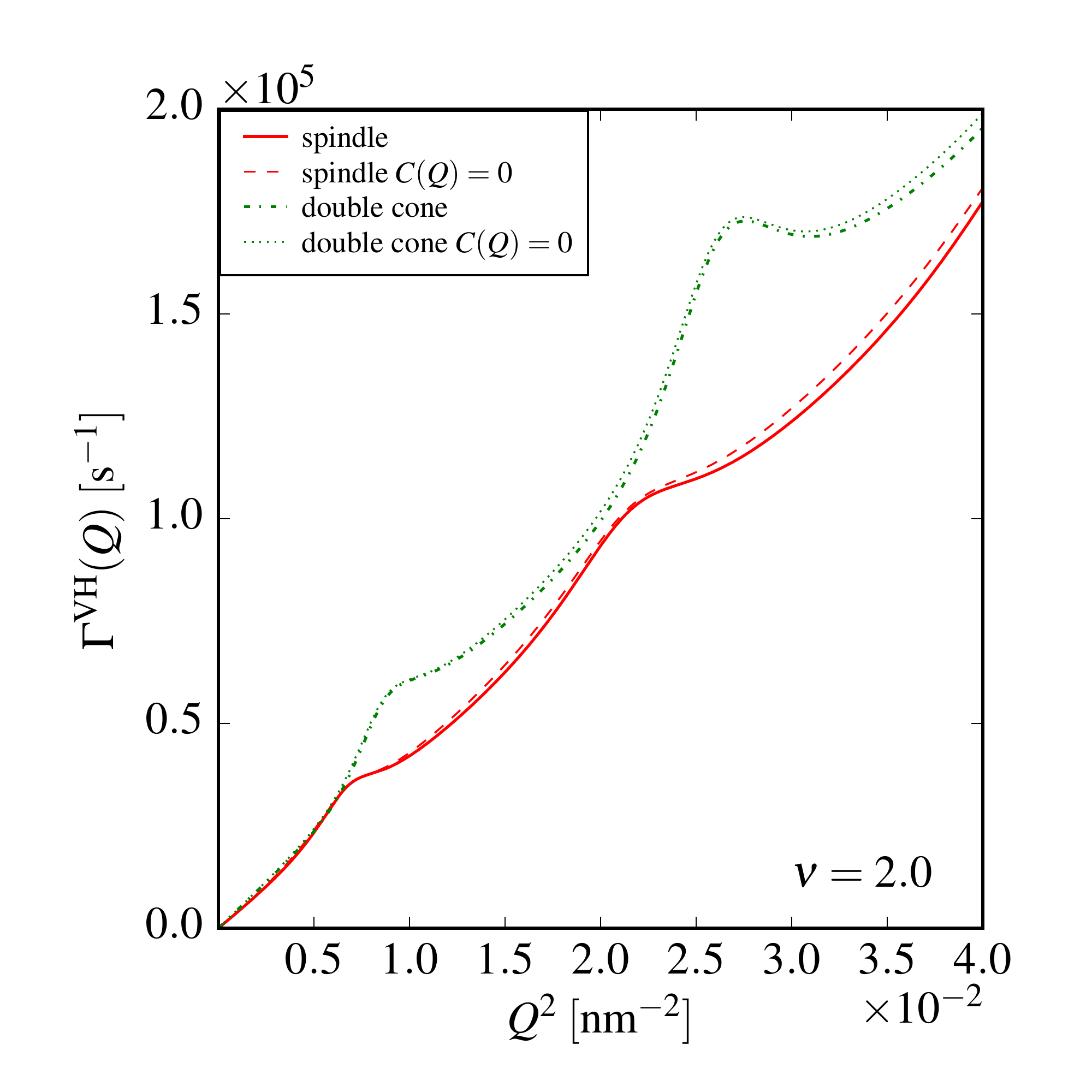}
\includegraphics[width=0.5\textwidth]{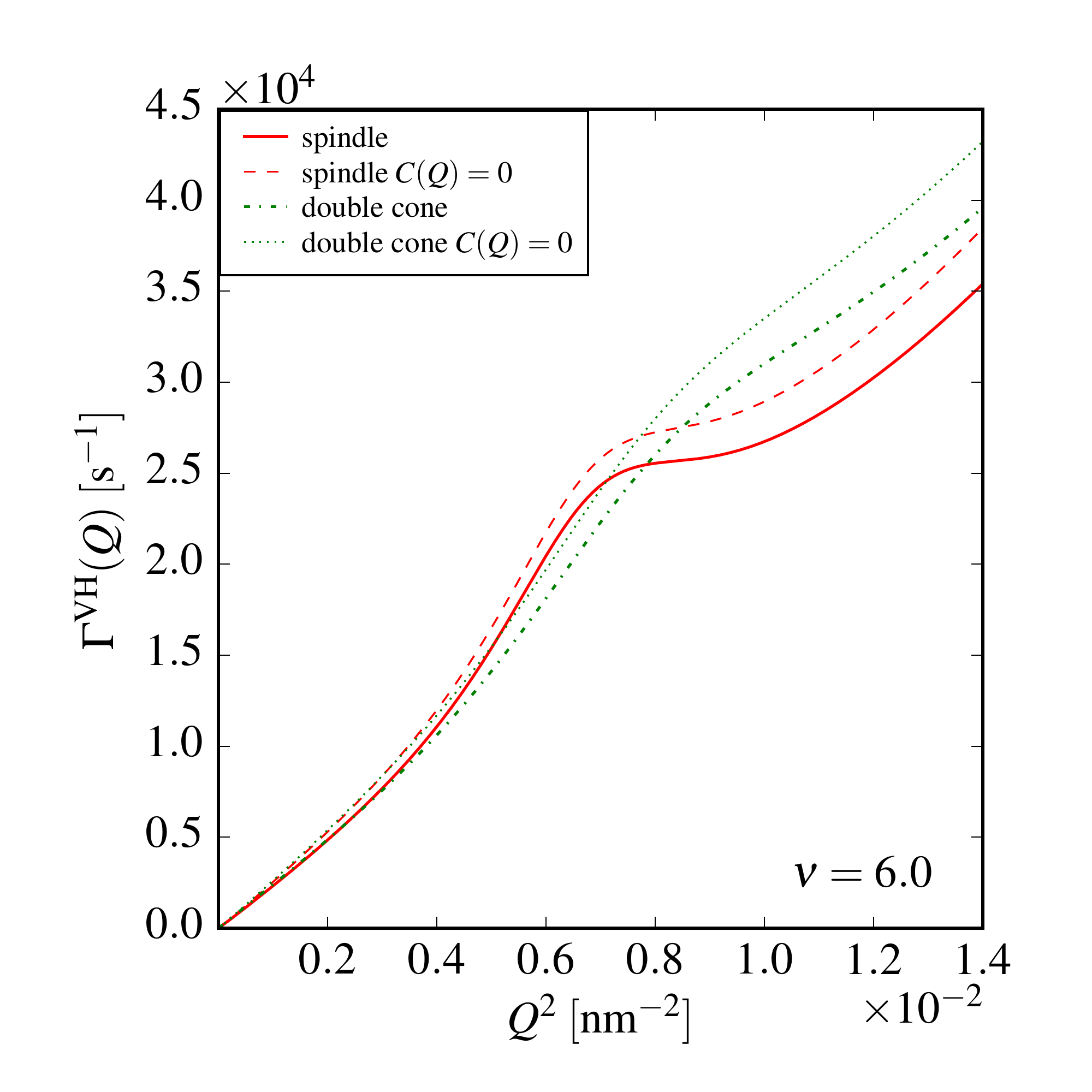}}
\caption{\label{fig:cumulants_without_coupling}First cumulants for differently shaped particles with aspect ratios of $\nu=2$ (lhs) and $\nu=6$ (rhs). The influence of the coupling function is shown by comparison of the first cumulants 
$\Gamma^{\rm VH}(Q)$ resulting from depolarized scattering with coupling function $C(Q)$ and neglecting the coupling function ($C(Q)=0$).}
\end{figure}

The influence of the coupling function can be seen in Fig.\ \ref{fig:cumulants_without_coupling}, where the first cumulants
$\Gamma^{\rm VH}(Q)$ are compared to the results neglecting the coupling function. The latter case is represented by the dashed lines.
For spindles and double cones as objects with pronounced apices, the influence of the translational-rotational coupling 
cannot be neglected at large scattering vectors. If the coupling function $C(Q)$ is negative, for prolate objects with $\Delta D_{\rm tr}>0$ the quadrupole contribution of the diffusion tensor leads to a slower relaxation of the depolarized scattering function. For cylinders and spherocylinders strong modulations of $C(Q)$ occur which lead to an accelerated relaxation if $C(Q)>0$.

\section{\label{conclusion}Conclusions}
The macroscopic rheological properties of fluids are considerably influenced by the mesoscale dynamics of suspended particles. For shape-anisotropic particles, in addition to translational diffusion, rotational diffusion is relevant for the rheological behavior, which depends not only on the volume fraction but 
also on the aspect ratio and shape of the particles.
  
Depolarized dynamic light scattering gives access to both translational and rotational diffusion of suspended particles. Even in highly dilute suspensions without particle interactions, the depolarized scattering function shows a nontrivial dependence on the scattering vector $Q$ if at least one typical particle dimension is
comparable to or larger than the wavelength. The first cumulants of the intermediate scattering function deviate from the Landau-Placzek behavior with a proportionality to $Q^2$. The modulation of the cumulants strongly depends on the particle shape and is related both to the rotational diffusion and the
translation-rotation coupling. The contribution of the coupling to the initial relaxation rates can either be positive or negative, depending on the particle geometry and the wave vector. For prolate objects, $\Delta D=D_\parallel - D_\perp$ is positive. For the calculation of $g_1^\mathrm{VH}(Q,t)$ a multipole expansion of 
the scattering function is required. With increasing aspect ratio of the particles an increasing number of expansion coefficients is needed to achieve a satisfactory convergence for an appropriate range of scattering vectors. However, in light scattering experiments, the maximum accessible scattering vector is limited to $Q_{\rm max}=4\pi n/\lambda$, where $\lambda$ is the wavelength and $n$ the refractive index of the suspending medium. With an effective wavelength of $\lambda/n\approx 400\,{\rm nm}$ for a green laser in aqueous suspensions of particles with $R_{\rm eq}=50\,{\rm nm}$, which is approximately the typical equatorial radius of spindle-shaped hematite particles,\cite{maerkert:2011} the maximum reduced scattering vector is roughly $QR_{\rm eq}\approx 1.5$.
Both translational and rotational diffusion coefficients of spindles and double cones are due to the smaller surface and volume of these objects significantly larger than those of spherocylinders and cylinders. As a consequence, also the difference $\Delta D_{\rm tr}$ is larger for the former shapes with pronounced apices.
Hence, the effect of the translation-rotation coupling visible in the first cumulants is for these particles larger than for the other ones. Hematite particles are a promising model system to investigate the coupling between rotational and translational diffusion. However, the size of these particles is too small to cover a sufficient $Q$-range employing visible light as a probe.
With coherent and linearly polarized X-rays from synchrotron sources and upcoming free electron laser sources, however, practically no limitation for the maximum 
scattering vector in the Q-range relevant for mesoscale objects exists. 
Employing  X-ray photon correlation spectroscopy with polarization analysis,\cite{Detlefs:2012} a sufficiently large range of scattering vectors can be accessed to address both translational and rotational diffusion of anisotropic particles. From the first cumulants determined over a sufficiently large  range of scattering vectors, the quantities $\langle D_{\rm tr}\rangle$, $\Delta D_{\rm tr}$, and $D_{\rm rot}$ relevant for the diffusion of anisotropic particles can be obtained experimentally.

Our approach can easily be extended to core-shell particles as demonstrated in Appendix B. In the case where an anisotropic core is embedded in a spherical shell, also systems with isotropic topology and anisotropic scattering power can be treated. Since for a spherical shell $\Delta D=D_\parallel-D_\perp$ vanishes, a
rotation-translation coupling is not present for such a system. Only the multipole expansion of the anisotropic scattering power modulates in this case the $Q^2$-proportionality of the first cumulants.

The approach can also be adapted to describe shape-anisotropic particles consisting of optically anisotropic materials with a tensorial polarizability. As long as the diffusion tensor is diagonal, also scattering functions for shapes without rotation and inversion symmetry can be calculated. In this case, additional expansion coefficients $f_{l,m\neq 0}$ have to be considered. Without optical inversion symmetry, also coefficients with odd order of $l$  are nonzero.

Moreover, for long rods, effects of particle
flexibility may get important. Fluctuations in the particle shape
can be incorporated into the present analysis
by considering a smeared averaged static form factor. The full
description of flexibility effects
beyond this simple effective picture, however, requires a nontrivial 
extension of our theory  which is left for future studies.

Our analysis provides a systematic framework to interpret depolarized light scattering data with respect to details in the shape of prolate particles. This is of great importance to the precise characterization of suspensions (beyond their aspect ratio) which contain anisotropic nano- or colloidal particles. One important example are ferrofluids which involve anisotropic magnetic particles in a carrier fluid and are of great importance to many magnetorheological applications.\cite{Odenbach_book}

Future work needs to be performed for oblate particles and for particles of arbitrary shape that are not rotationally symmetric. The latter case requires a more 
sophisticated description for the orientations (e.g., by using Eulerian angles) such that the Brownian motion is much more complicated.\cite{Wittkowski:2012,Kraft:13}
Another line of future research is to incorporate the correlations at finite particle density. Particularly interesting is the high-density regime where the suspension is getting crystalline \cite{Pal:2015} or glassy. An explicit computation of the scattering function will, however, become much more complicated than the analysis presented here due to the intricate character of the orientation-dependent interactions. Still, in phases with strict periodic order, such as fully
crystalline phases or plastic crystals, one can hope that the analysis is somewhat easier since one can apply natural periodic boundary conditions. An important question is whether one can achieve a quantitative level of understanding of the role of translational and rotational fluctuations of the particles in the stability of specific ordered phases.
Finally, an external field can be applied which couples to the orientational degrees of freedom and gives rise to new phase phenomena, such as a paranematic-nematic transition,\cite{GrafJPCM:1999} new kinds of crystals,\cite{Demiroers:2010} and an introduction or destruction of biaxial orientational order (see, e.g., Refs.\ \citenum{Shimbo:2006} and \citenum{Ghoshal:2014}). An extension of our analysis to the case of noninteracting particles in an orienting external field should be feasible and will be a topic of our future research.

\appendix
\section{Meridian curves for solids of revolution}

The meridian curve for a cylinder simply is $x(z)=1$. For $r\le\nu$ only one symmetrically independent intersection with a 
sphere exists, for $\nu < r \le \sqrt{\nu^2+1}$ an additional circle exists as intersection. The critical pole distances are
\begin{align}
\xi=\cos\vartheta_c &=\left\{ \begin{array}{ll}
\pm\left(\dfrac{r^2-1}{r^2}\right)^\frac{1}{2} & r\le\nu \\
\left\{ \begin{array}{l}
\pm\left(\dfrac{r^2-1}{r^2}\right)^\frac{1}{2} \\
\pm\left(1-\dfrac{r^2-\nu^2}{r^2}\right)^\frac{1}{2}
\end{array}\right\}& \nu<r\le\sqrt{\nu^2+1} \,.
\end{array}
\right.
\end{align}

In the region $\nu<r \le \sqrt{\nu^2+1}$ the two symmetrically independent intersections lead to the 
expansion coefficients
\begin{align}
f_{2l}(r) &=2\sqrt{\left(4l+1\right)\pi}\left[\Xi_{2l}(\xi_1(r))-\Xi_{2l}(\xi_2(r))\right].
\end{align} Here, $|\xi_1(r)|<|\xi_2(r)|$ denote the cosines of the two independent critical pole distances.

The meridian curve of an ellipsoid is given by 
\begin{align}
x(z)&=\left( 1- \dfrac{z^2}{\nu^2}\right)^\frac{1}{2}.
\end{align}
Only one symmetrically independent intersection with a sphere at the critical pole distance
\begin{align}
\xi &=\cos\vartheta_c=\left(\dfrac{r^2-1}{\nu^2-1}\right)^{\frac{1}{2}}
\end{align}
exists.

A spindle with the equatorial radius $R_{\rm eq}$ and the length $2\nu R_{\rm eq}$ can be described as a solid of revolution resulting from the intersection of two circles with the radius $R_{\rm eq}(\nu^2+1)/2$, the centers of which have a distance $R_{\rm eq}(\nu^2-1)$. For the volume the analytical expression 
\begin{align}
V=\dfrac{2\pi}{3}\left(R_{\rm eq}^3\nu\left(\frac{3}{4}\left(1+\nu^2\right)^2-\nu^2\right)+\frac{3R_{\rm eq}} {4}\left(1+\nu^2\right)^2\left(1-\frac{1}{2}\left(1+\nu^2\right)\right) \arcsin \left(\frac{2\nu}{1+\nu^2}\right)\right)
\end{align}
is obtained.\cite{maerkert:2011} The corresponding meridian curve
\begin{align}
x(z)=1-\dfrac{\nu^2+1}{2}+\dfrac{1}{2}\left(\nu^2(\nu^2+2)-4z^2+1\right)^{\frac{1}{2}} 
\end{align}
leads to one symmetrically independent critical pole distance 
\begin{align}
\xi=\cos\vartheta_c=\pm\left(1-\left(\frac{r^2-\nu^2}{r\left(1-\nu^2\right)}\right)^2\right)^{\frac{1}{2}} .
\end{align}

\begin{figure}[t]
\centerline{\includegraphics[width=0.45\textwidth]{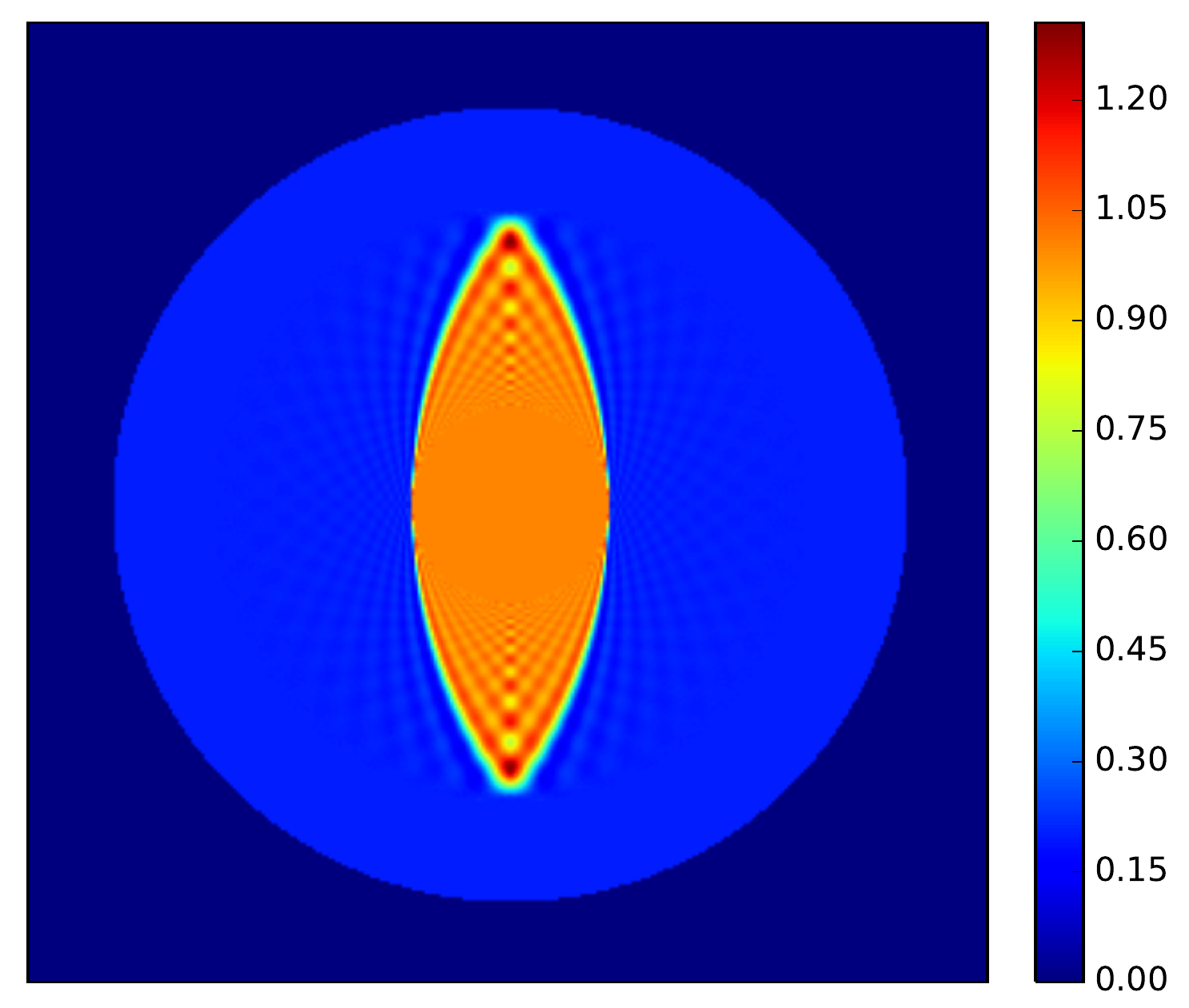}}
\caption{\label{fig:core_shell_reconstruction} Reconstruction of the meridional cross section of a core-shell structure consisting of a spindle-shaped core with a length $L=0.75 \sigma_s$ and an aspect ratio of $\nu=3$ in a spherical shell with the outer diameter $\sigma_s$. The scattering length density of the core is $\rho_c=1$ and that of the shell $\rho_s=0.2$.}
\end{figure}

\begin{figure}[ht]
\centerline{
\includegraphics[width=0.5\textwidth]{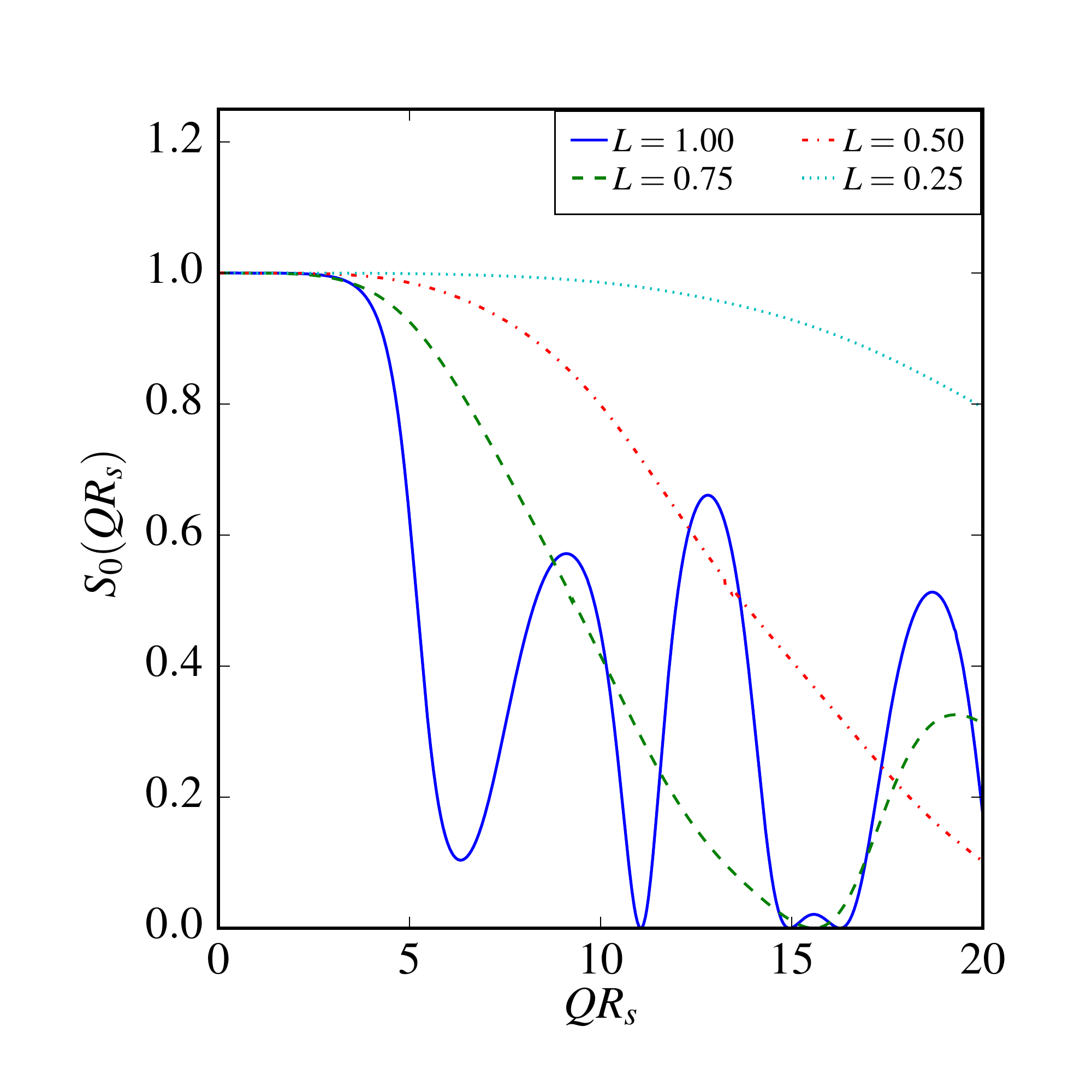}
\includegraphics[width=0.5\textwidth]{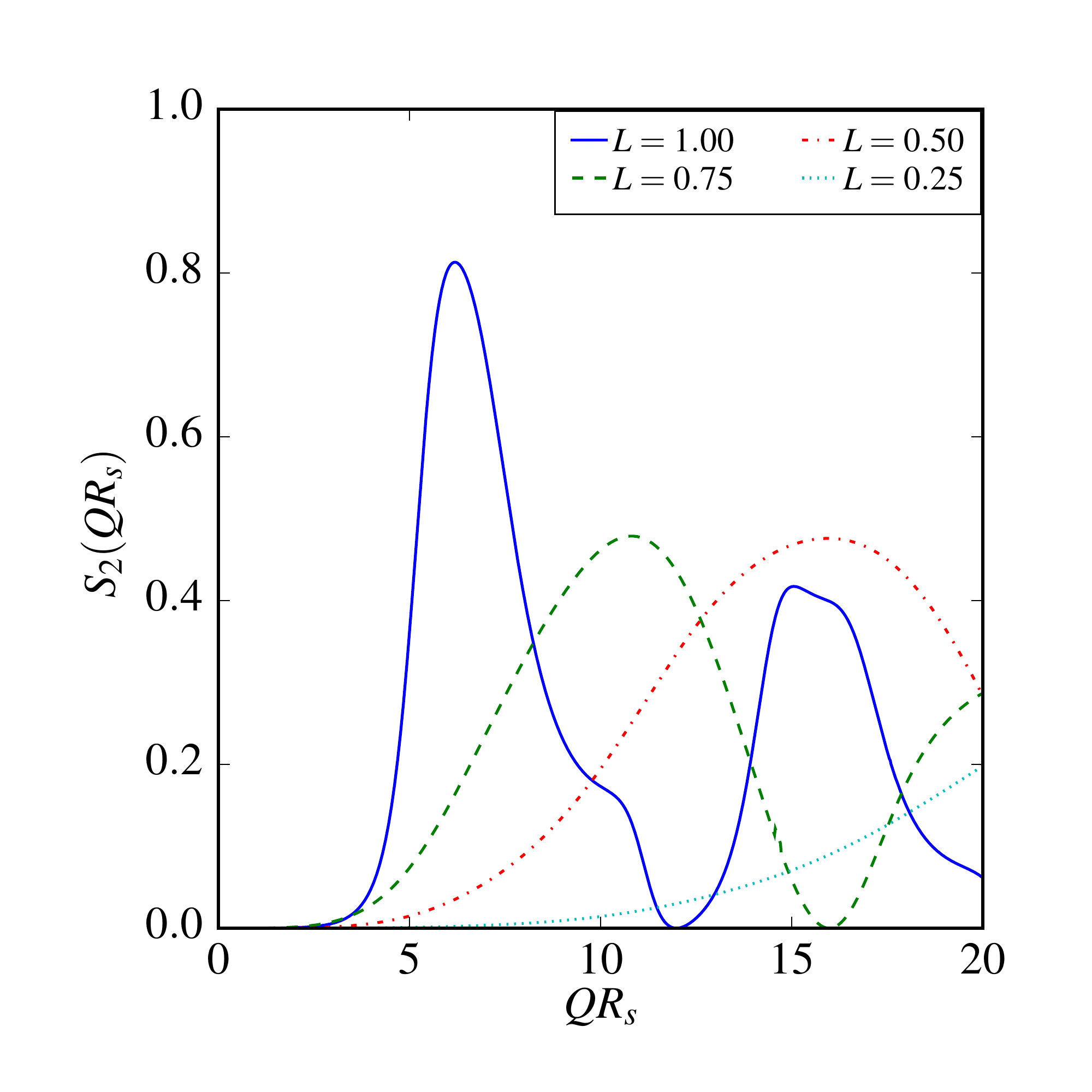}}
\centerline{
\includegraphics[width=0.5\textwidth]{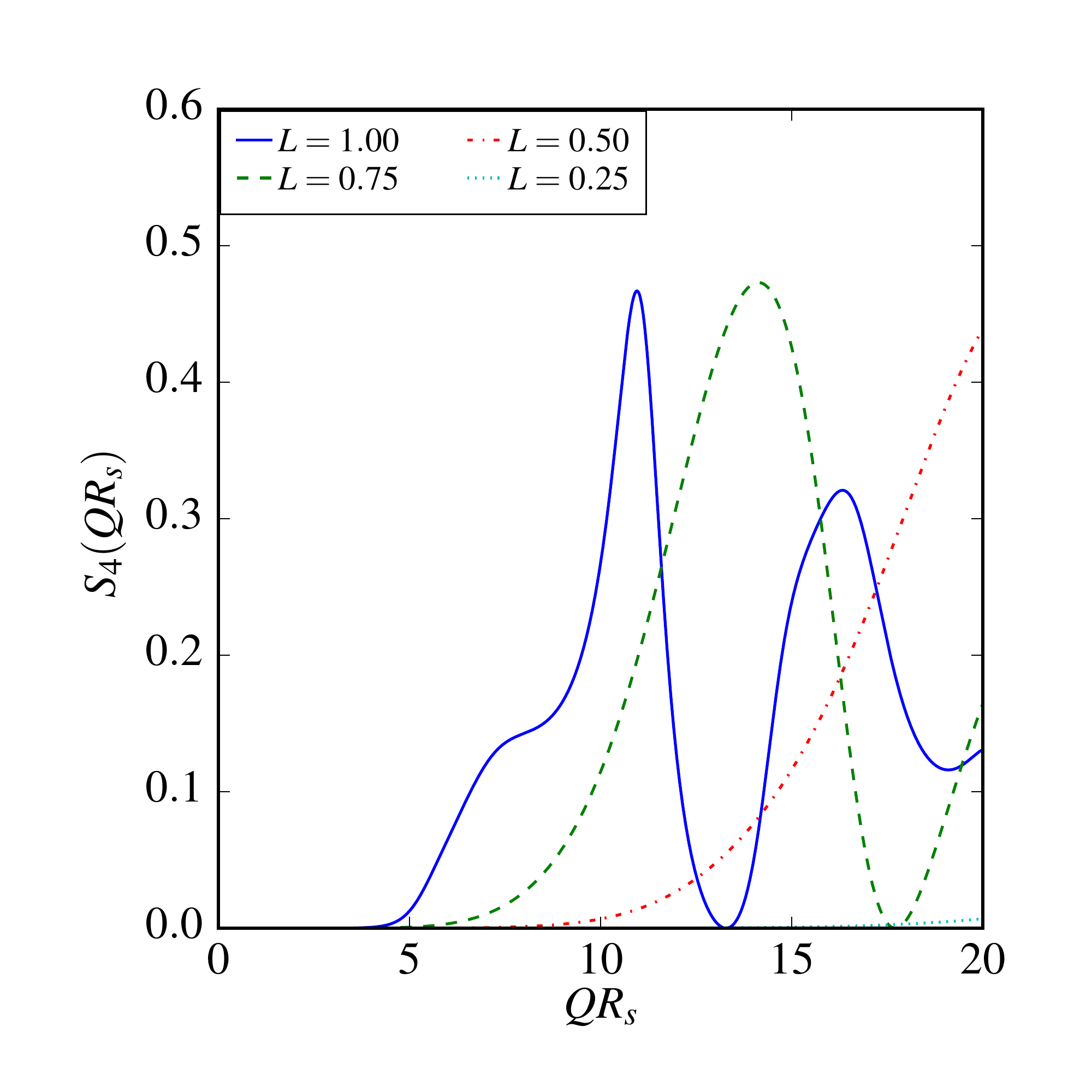}
\includegraphics[width=0.5\textwidth]{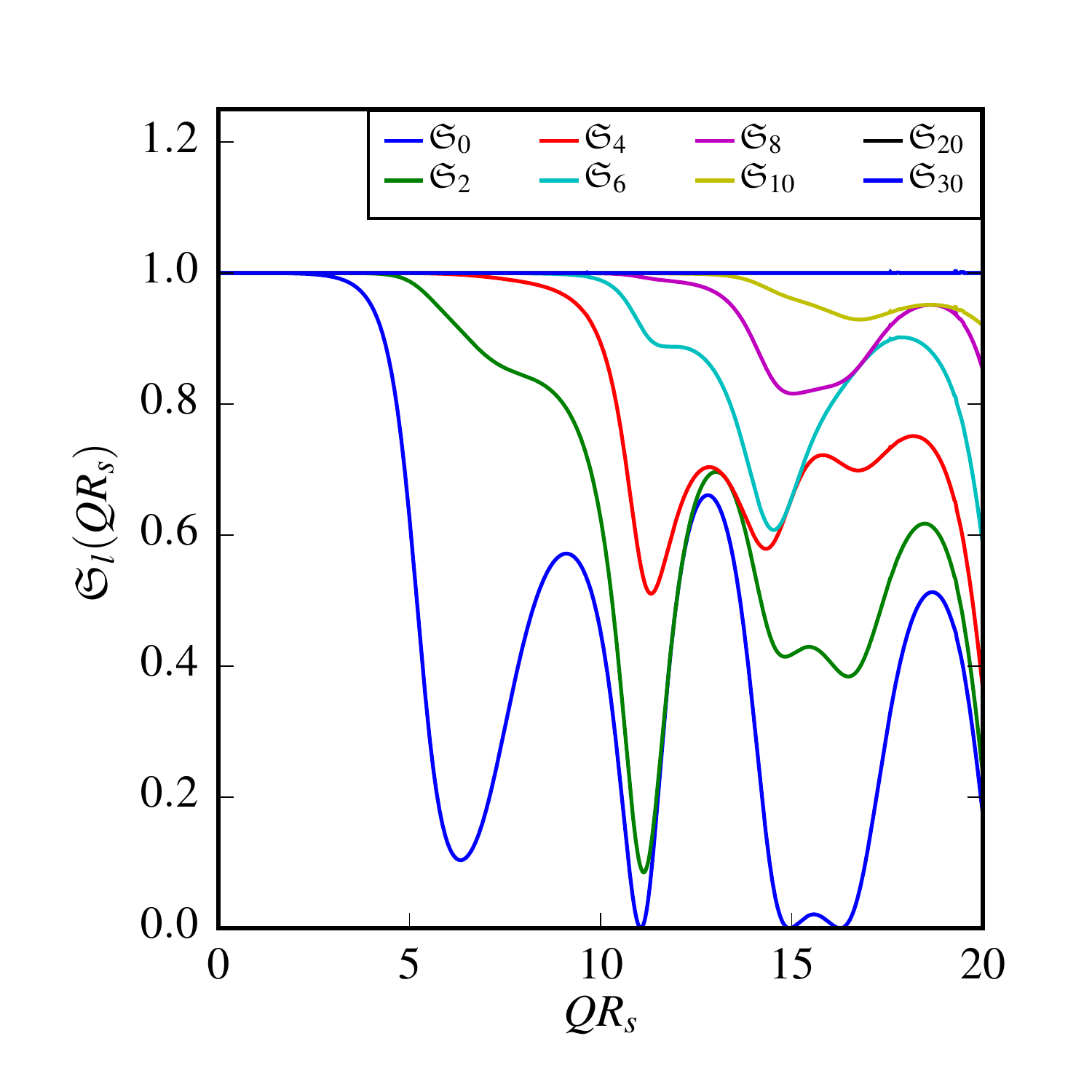}}
\caption{\label{fig:core_shell_expansion}
Expansion coefficients $S_{2l}(Q)$ for core-shell particles consisting of a spindle with the aspect ratio $\nu=3$ embedded in a spherical shell for different lengths $L/\sigma_s$ of the spindles. In the lower right corner, exemplarily a convergence test of the expansion for $L/\sigma_s=1$ is displayed.}
\end{figure}

\begin{figure}[ht]
\centerline{\includegraphics[width=0.5\columnwidth]{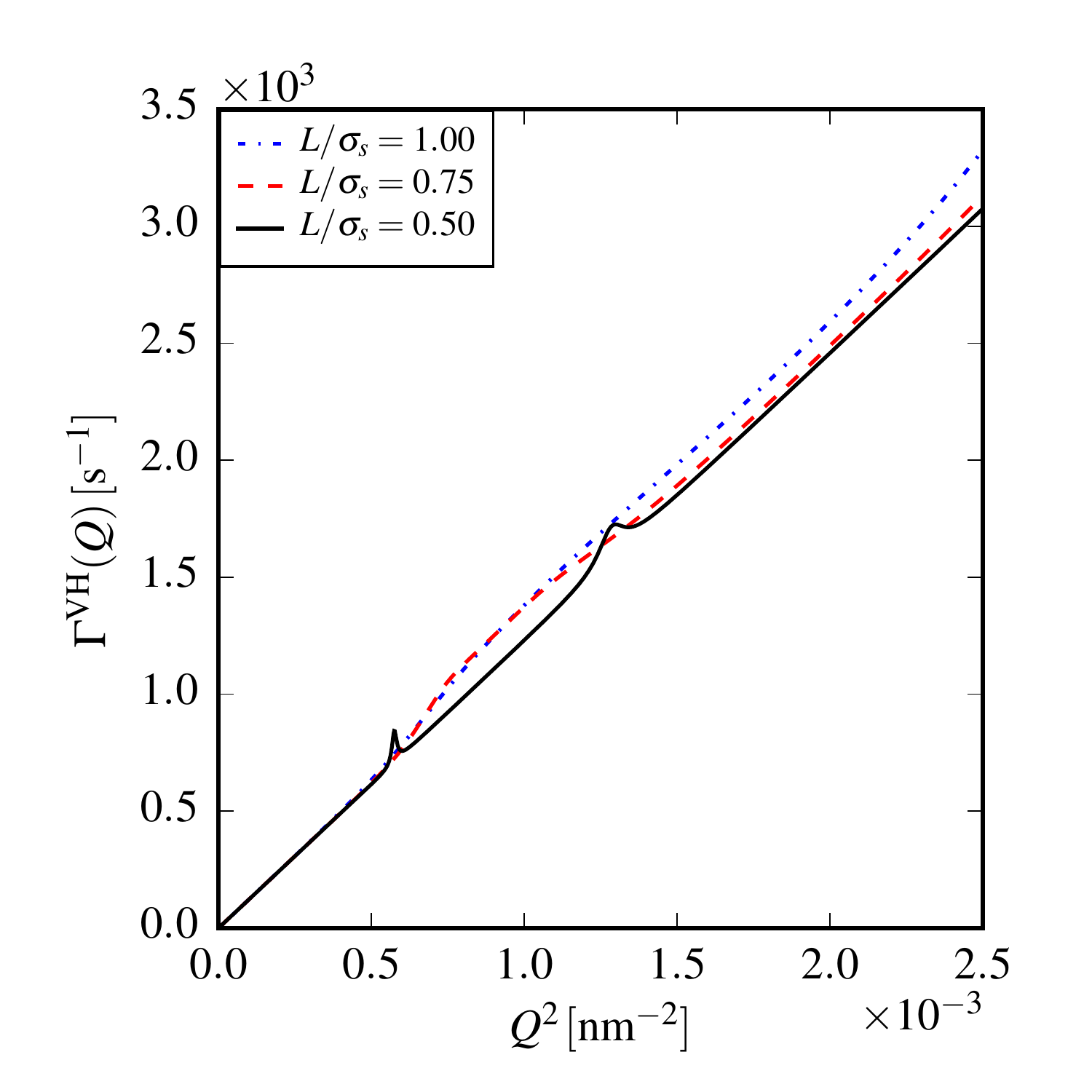}}
\caption{\label{fig:cumulants_core_shell}First cumulants for core-shell particles consisting of a spindle-shaped core with an aspect ratio of $\nu=3$ and a scattering length density of $\varrho_s=1$ and a spherical shell with an hydrodynamic radius of $R_s=200\,{\rm nm}$ and a scattering length density of $\varrho_s=0.2$ for different sizes of the anisotropic core. The relative size of the core is quantified by the ratio of the length $L$ of the spindle and the hydrodynamic diameter $\sigma_s$ of the spherical shell.}
\end{figure}

\section{Core-shell particles}

The approach of expanding form factors in terms of spherical harmonics can be extended to core-shell structures. In the special case where an anisotropic core is embedded into a spherical shell, optically anisotropic structures undergoing isotropic diffusion processes can be described.
The translational and rotational diffusion coefficients determined by the spherical shape of the shell can be calculated employing the Stokes-Einstein and Stokes-Einstein-Debye relations. Since the translational diffusion tensor is spherically symmetric, with $\Delta D=0$ the coupling function is irrelevant for such a core-shell structure. The formal multipole expansion of the anisotropic core's scattering power, however, leads to a modulation of the $Q^2$-dependence of the first cumulants.

In Fig.\ \ref{fig:core_shell_reconstruction} the meridional cross section of a core-shell structure reconstructed from the expansion in 
spherical harmonics is displayed. Here a spindle-shaped core with the scattering length density $\rho_c=1$ and an aspect ratio of $\nu=3$ is embedded in a spherical shell with a scattering length density $\rho_s=0.2$.  The length of the spindle is $L=2\nu R_{\rm eq}=0.8 \sigma_s$, where $\sigma_s=2R_s$ denotes the outer diameter of the spherical shell and $R_{\rm eq}$ the equatorial radius of the core.

The expansion coefficients for core-shell systems with differently sized anisotropic cores are displayed in Fig.\ \ref{fig:core_shell_expansion}. The smaller the size of the anisotropic core is compared to the spherical shell, the less important are the multipole contributions to the scattering power. Although the coupling function is irrelevant for these systems, depending on the relative size of the anisotropic core, in the first cumulants
deviations from the proportionality to $Q^2$ occur, which originate from the $Q-$dependent multipole contributions to the scattering power. In Fig.\ \ref{fig:cumulants_core_shell}, the first cumulants for core-shell particles with different sizes of the anisotropic core are displayed. The radius of the spherical shell is assumed to be $R_s=200\,{\rm nm}$. The translational and rotational diffusion coefficients are calculated by means of the Stokes-Einstein and Stokes-Einstein-Debye relations using the viscosity $\eta=8.9\times 10^{-4}\,{\rm Pa\,s}$ of water at 298\,K.

\begin{acknowledgments}
The authors thank J.~K.~G.~Dhont for fruitful discussions.
This work was supported by the Deutsche Forschungsgemeinschaft (DFG) through the priority program SPP 1681 on ferrogels. J. W. acknowledges financial support by the BMBF within the R\"ontgen-{\AA}ngstr{\o}m cluster.
\end{acknowledgments}


\bibliography{refs}

\end{document}